\documentclass{aa}  

\usepackage{graphicx}
\usepackage{natbib}
\usepackage{txfonts}
\usepackage{caption}
\usepackage{subcaption}
\usepackage{appendix}
\usepackage{rotating}
\usepackage{longtable}
\usepackage{lscape}
\begin{document}

    \title{Experimental and theoretical oscillator strengths of \ion{Mg}{I} for accurate abundance analysis}

    \author{A. Pehlivan Rhodin
          \inst{1,2}
          \and
          H. Hartman\inst{1,2}
          \and
H. Nilsson\inst{2}
          \and
P. J\"{o}nsson\inst{1}
          }

   \institute{Materials Science and Applied Mathematics, Malm\"{o} University, 205 06 Malm\"{o}, Sweden \\
             \email{asli.pehlivan@mah.se, asli@astro.lu.se} 
   \and
         Lund Observatory, Box 43, SE-221 00 Lund, Sweden\\
             }
   \date{Received 10 October, 2016; accepted 18 November, 2016}

 
  \abstract
   {With the aid of stellar abundance analysis, it is possible to study the galactic formation and evolution. Magnesium is an important element to trace the $\alpha$-element evolution in our Galaxy. For chemical abundance analysis, such as magnesium abundance, accurate and complete atomic data are essential. Inaccurate atomic data lead to uncertain abundances and prevent discrimination between different evolution models.}
   {We study the spectrum of neutral magnesium from laboratory measurements and theoretical calculations. Our aim is to improve the oscillator strengths ($f$-values) of \ion{Mg}{I} lines and to create a complete set of accurate atomic data, particularly for the near-IR region.   }
   {We derived oscillator strengths by combining the experimental branching fractions with radiative lifetimes reported in the literature and computed in this work. A hollow cathode discharge lamp was used to produce free atoms in the plasma and a Fourier transform spectrometer recorded the intensity-calibrated high-resolution spectra. In addition, we performed theoretical calculations using the multiconfiguration Hartree-Fock program ATSP2K.  }
    {This project provides a set of experimental and theoretical oscillator strengths. We derived 34 experimental oscillator strengths. Except from the  \ion{Mg}{I} optical triplet lines (3p $^3$P$^o_{0,1,2}$ - 4s $^3$S$_1$), these oscillator strengths are measured for the first time. The theoretical oscillator strengths are in very good agreement with the experimental data and complement the missing transitions of the experimental data up to $n=7$ from even and odd parity terms.  We present an evaluated set of oscillator strengths, \textit{gf}, with uncertainties as small as 5\%. The new values of the \ion{Mg}{I} optical triplet line (3p $^3$P$^o_{0,1,2}$ - 4s $^3$S$_1$) oscillator strength values are $\sim$0.08 dex larger than the previous measurements.}
   {}

   \maketitle
%

\section{Introduction}

Magnesium is an important element for chemical evolution studies. It is an $\alpha$-element, which is formed and released during supernova type II explosions of massive stars. Magnesium lines are strong in the spectra of late-type stars and even in metal-poor stars. Therefore, it is an ideal element to trace the $\alpha$-element abundances. \\
\indent The dominant electron source in the stellar atmospheres of metal-poor stars is magnesium. As a result, its abundance affects the model atmospheres \citep{Prochaska}. The higher the magnesium abundance, the higher the electron density becomes in the stellar atmosphere. Neglecting this fact may lead to incorrect stellar gravity determination.  \citet{Prochaska} used an $\alpha$-enhanced model atmosphere to derive abundances. For magnesium abundance analysis, they only found very few magnesium lines with reported $\log(gf)$ values. Because of the missing data, they included additional lines with astrophysical $\log(gf)$ values.\\
\indent Several studies \citep{Shigeyama, Bensby, Cayrel,Andrievsky} have used magnesium as an alternative to iron for tracing the chemical evolution of the Milky Way. Magnesium is only formed in supernova type II explosions of massive stars \citep{Woosley}, whereas iron has several formation channels \citep{Thielemann}. A complete set of magnesium atomic data results in more accurate abundances and, correspondingly, makes magnesium an even better choice as a tracer of galactic evolution.  \\
\indent At temperatures $T \geq 5000K$, magnesium is primarily singly ionised. However there are a large number of \ion{Mg}{I} lines existing in the solar spectrum \citep{Scott}. As a result of Mg$^+$ being the dominant species, \ion{Mg}{I} is sensitive to the deviations from local thermodynamic equilibrium (LTE). In particular, for the metal-poor stars, these non-LTE effects are predicted to be significant \citep{Zhao1, Zhao2}. To study the deviations from LTE, It is crucial to have accurate atomic data of both \ion{Mg}{I} and \ion{Mg}{II}. This makes it possible to map the limits of LTE approximations as a function of stellar metallicity, gravity, and temperature, similar to \ion{Fe}{I} in \citet{Lind}. There are several studies on NLTE analysis of neutral magnesium including the recent studies of \citet{Bergemann, Osorio}.The former studied NLTE effects in the J-band \ion{Mg}{I} lines and due to a lack of experimental $\log(gf)$ values, calculated $\log(gf)$ values were used. However, using the average of the many calculated $\log(gf)$ values overestimated the line depths, \citet{Bergemann} concluded that the values were wrong and derived their astrophysical $\log(gf)$ values.\\
\indent \citet{Scott} determined the magnesium abundance of the Sun to be $\log \epsilon_{\text{Mg}} =  7.59 \pm0.04$ from a 3D hydrodynamic model of the solar photosphere. However, due to the lack of laboratory measurements of $\log(gf)$ values, they used theoretical $\log(gf)$ values of \citet{Butler} and \citet{Chang}. The current study provides experimental $\log(gf)$ values for two of the lines and improved theoretical $\log(gf)$ values for all the lines used by \citet{Scott}. \\
\indent In addition, some planetary atmosphere studies show the presence of magnesium in the atmospheres of planets \citep{Fossati, Vidal, Bourrier1, Bourrier2}. By analysing the resonance line of an abundant element, such as magnesium, during a planet transit, the atmospheric escape mechanism can be understood. These studies are usually done by analysing absorption depths of the line of interest, which requires accurate atomic data. \\
\indent To our knowledge there are no experimental oscillator strengths of \ion{Mg}{I} lines, except from the $3\text{s}^2 $ ${}^1\text{S}_0 - 3\text{s}3\text{p}$ $ {}^3\text{P}_1$ intercombination transition at 4571 $\AA$  \citep{Kwong} and the \ion{Mg}{I} triplets lines (3p $^3$P$^o_{0,1,2}$ - 4s $^3$S$_1$); 5167, 5172, and 5183 $\AA$  \citep{Aldenius}. Although \citet{Ueda} provided oscillator strengths for the transitions from 3p $^3$P level, they were not completely experimental values. Theoretical calculations in \citet{Wiese} compilation were used for the absolute scale of the oscillator strengths. \\
\indent There are several theoretical values, which are generally used for abundance analysis. \citet{Chang} calculated oscillator strengths of \ion{Mg}{I} lines between selected $^{1,3}$S - $^{1,3}$F states using the configuration interaction (CI) procedure with a finite basis set constructed from \textit{B splines}. In addition, theoretical values of \citet{Butler} are commonly used for abundance analyses. They used the close-coupling approximation with the R-matrix technique. Moreover, \citet{Civis} performed oscillator strength calculations using the quantum defect theory (QDT) in the region of $800-9000$ cm$^{-1}$. \citet{Froese} performed calculations using the multi configuration Hartree-Fock method. Their calculations included the terms up to $n=4$ and all three types of correlations: valence, core-valence, and core-core correlation.    \\
\indent This paper presents experimental $\log(gf)$ values of \ion{Mg}{I} lines from high-resolution laboratory measurements in the infrared and optical region from the upper even parity 4s $^{1,3}$S, 5s $^1$S, 3d $^1$D, and 4d $^1$D terms and the odd parity 4p $^3$P$^o$, 5p $^3$P$^o$, 4f $^{1,3}$F$^o$, and 5f $^{1,3}$F$^o$ terms. In addition, we performed multiconfiguration Hartree-Fock calculations using the ASTP2K package \citep{Charlotte} and obtained $\log(gf)$ values of \ion{Mg}{I} lines up to $n=7$ from even parity $^{1,3}$S, $^{1,3}$D, and$^{1,3}$G terms and odd parity $^{1,3}$P$^o$, and $^{1,3}$F$^o$ terms. The transitions between the higher terms fall in the IR spectral region and the calculated $\log(gf)$ values are important for interpreting observations using the new generation of telescopes designed for this region. Following the introduction, Sect. 2 describes the experimental method we used for deriving $\log(gf)$ values. In addition, this section explains the measurements of branching fractions ($BF$) and the uncertainty estimations. The theoretical calculations that we performed are explained in Sect. 3. In Sect. 4, we present our results, the comparisons of our results with previous studies, and the conclusions. 
\section{Experimental method} 
We used a water-cooled hollow cathode discharge lamp (HCL) with a magnesium cathode as a light source to produce the magnesium plasma. The experimental set-up was similar to the one described by \citet{Pehlivan}. The strongest lines for the measurements were obtained using neon as carrier gas and with an applied current of $0.60$ A. \\
\indent We recorded the \ion{Mg}{I} spectra with the high-resolution Fourier transform spectrometer (FTS), Bruker IFS 125 HR, at the Lund Observatory (Edl{\'e}n Laboratory). The maximum resolving power of the instrument is $10^6$ at 2000 cm$^{-1}$ and the covered wavenumber region is $50000-2000$ cm$^{-1}$ ($200-5000$ nm). We set the resolution to $0.01$ cm$^{-1}$ during the measurements and recorded the spectra with indium antimonide (InSb), silicon (Si), and photomultiplier tube (PMT) detectors. These detectors are sensitive to different spectral regions, but they overlap each other in a small wavelength region. \\
\indent The optical element contributions to the FTS response function were compensated for by obtaining an intensity calibration. Because of the wavelength-dependent transmission of the optical elements and the spectrometer, the measured intensities of the lines differ from their intrinsic intensities. Therefore, we acquired the response function of the instrument for three different detectors that we used during different measurements. The response function is usually determined by measuring the spectrum of an intensity calibrated reference lamp. We used a tungsten filament lamp for the intensity calibration of \ion{Mg}{I} lines. The lamp was calibrated by the Swedish National Laboratory (SP) for spectral radiance in the region between $40 000-4000$ cm$^{-1}$ ($250-2500$ nm). With the calibrated radiance of the lamp, the response function of the instrument can be determined for different detectors. We used the overlapping region \ion{Mg}{I} lines, which were recorded with different detectors, to connect the relative intensities on the same scale. This was done by using a normalisation factor $nf$, which in turn contributed an additional uncertainty to the $BF$s.  \\
\indent In addition, we recorded the spectra with different currents to compensate for self-absorption effects. The self-absorption affects the intensity of the line and this, in turn, influences the \textit{BF} measurements which are used to determine the oscillator strengths. More details can be found in our previous paper \citep{Pehlivan}.

\subsection{Branching fraction measurements}
The oscillator strength of a spectral line is proportional to the transition probability. For electric dipole transition, it is given as
\begin{equation}\label{Eq1}
f=\frac{g_\text{u}}{g_\text{l}} \lambda^2 A_{\text{ul}} 1.499 \times 10^{-16},
\end{equation}
where $g_\text{u}$ is the statistical weight of the upper level, $g_\text{l}$ the statistical weight of the lower level, $\lambda$ the wavelength of the transition in {\AA}, and $A_{\text{ul}}$ the transition probability between the upper level u and the lower level l in s$^{-1}$.\\
\indent The radiative lifetime of an upper level, $\tau_\text{u}$ is the inverse of the sum of all transition probabilities from the same upper level, $\tau_{\text{u}}= 1/ \sum_i A_{\text{u}i}$. The branching fraction \textit{(BF)} of a line is defined as the transition probability of the line $A_{\text{ul}}$ divided by the total transition probability of the lines from the same upper level;
\begin{equation}\label{Eq2}
BF_{\text{ul}}= \frac{A_{\text{ul}}}{\sum_{\rm i} A_{\text{u}i}}= \frac{I_{\text{ul}}}{\sum_{\rm i} I_{\text{u}i}}.
\end{equation}
 As the transition probability is proportional to the line intensity $I_{\text{ul}}$, $BF$ can be defined as the ratio of the line intensities. \\
\indent Knowing the radiative lifetime and combining this with the measured $BF$s, one can derive the transition probability, $A_{\text{ul}}$, of a spectral line;
\begin{equation}\label{Eq3}
A_{\text{ul}}=\frac{BF_{\text{ul}}}{\tau_\text{u}}.
\end{equation}
\indent Transitions from the same upper level can have wavelengths belonging to different regions of the electromagnetic spectrum. However, to accurately measure $BF$s, all transitions from the same upper level should be accounted for. For this reason, we recorded \ion{Mg}{I} spectra using different detectors. These different spectra were put on the same relative intensity scale by using a normalisation factor.   \\
\begin{figure*}
\centering
\includegraphics[width=\textwidth]{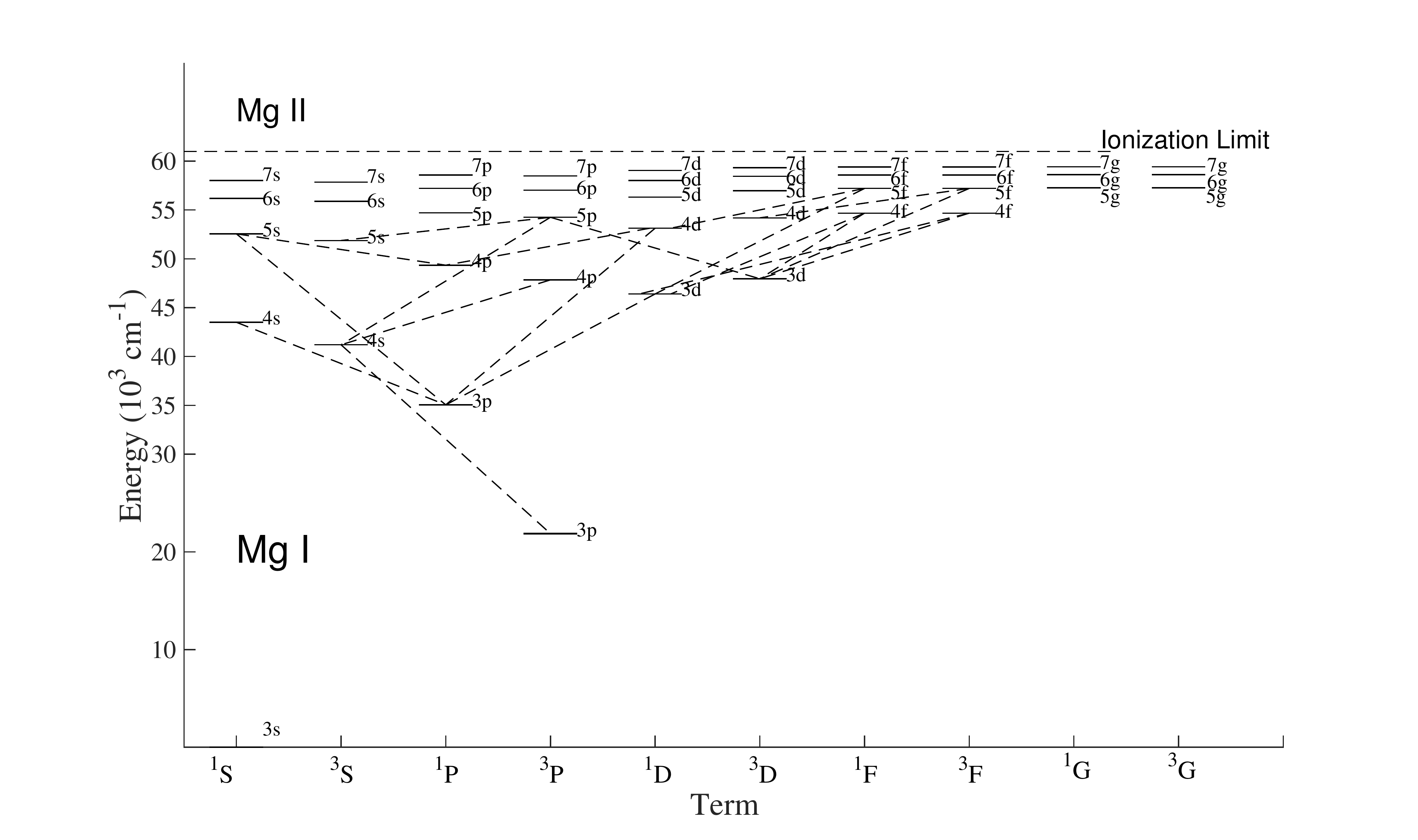}
\caption{Partial energy level diagram of \ion{Mg}{I} with dashed lines showing the observed transitions. The energy level values are from \citet{Martin}.}
\label{Fig1}
\end{figure*}
\indent A partial energy level diagram of \ion{Mg}{I} levels is shown in Fig. \ref{Fig1}. The transitions, which we observed and used to derive $\log(gf)$ values, are marked in this figure. Using the \citet{Kurucz} database and references in \citet{Kaufman}, we predicted the \ion{Mg}{I} lines from the same upper level. We identified these lines and analysed our recorded spectra with the FTS analysis software GFit \citep{Engstrom1, Engstrom2}.\\
\indent \ion{Mg}{I} has three dominant isotopes: \element[ ][24]{Mg} with 78.99\% abundance, \element[][25]{Mg} with 10\% abundance, and \element[][26]{Mg} with 11\% abundance (IUPAC 1991). Although there are three isotopes of \ion{Mg}{I}, in our measurements we did not see any isotope shift. The nuclear spins of these isotopes are $0$, $5/2$, and $0$, respectively. This proves that the most dominant isotope \element[][24]{Mg} has no hyperfine splitting (hfs) in the line profiles. Even though  \element[][25]{Mg} has a nuclear spin of $5/2$, we did not see any hfs as the abundance of this isotope is very low compared to \element[][24]{Mg}. \\

\subsection{Uncertainties}
\begin{figure}
\centering
\includegraphics[width=0.5\textwidth]{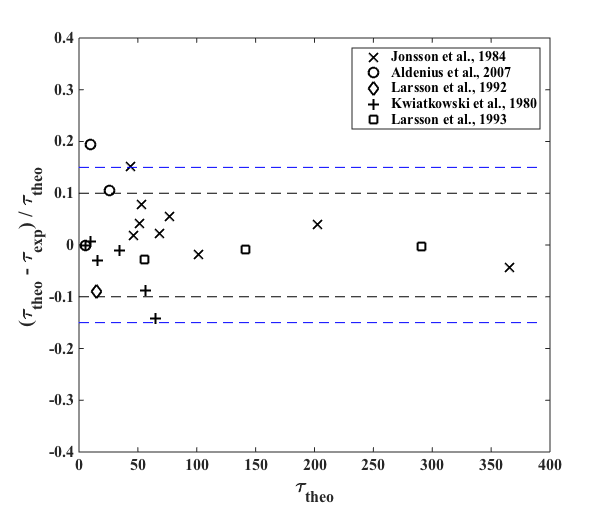}
\caption{Comparison of the theoretical lifetimes of this study with the previously measured experimental lifetimes. As seen in the figure, there is a large difference of almost $20\%$ for one of the values measured by \citet{Aldenius}. However, a re-measurement of the lifetime brings it in a very good agreement with the calculated value (see text for more details).}
\label{Fig2}
\end{figure}

The uncertainty of the $BF$ contains several components. Together with the uncertainty of the intensities, the uncertainty of the self-absorption correction, the uncertainty of the intensity calibration lamp, and the uncertainty of the normalisation factor, which is used to put the intensities on the same scale, should be considered. Including all of these uncertainty components, \citet{Sikstrom} defined the total uncertainty of the $BF$ as,
\begin{align} 
\Bigg(\frac{u(BF)}{BF} \Bigg)^2 &= (1-(BF)_k)^2 \hspace{0.6mm}\Bigg(\frac{u(I_k)}{I_k} \Bigg)^2  \nonumber \\
&+ \sum\limits_{j \neq k(inP)} (BF)_j^2 \hspace{0.8mm} \Bigg( \bigg( \frac {u(I_j)}{I_j} \bigg )^2 + \bigg( \frac{u(c_j)}{c_j} \bigg)^2 \Bigg) \nonumber \\
&+ \sum\limits_{j \neq k(inQ)} (BF)_j^2 \hspace{0.8mm} \Bigg( \bigg( \frac {u(I_j)}{I_j} \bigg )^2 + \bigg( \frac{u(c_j)}{c_j} \bigg)^2 +\bigg(\frac{u(nf)}{nf} \bigg)^2 \Bigg).
\end{align}
The first term of the equation includes the branching fraction $(BF)_k$ of the line of interest in the spectral region of the detector $P$ and the uncertainty in the measured intensity of the same line, $u(I_k)$. In the sum that follows $u(c_j)$  and $u(I_j)$ are the uncertainties of the calibration lamp and the uncertainties of the measured intensities, respectively, for other lines from the same upper level recorded with the detector P. $(BF)_j$ are the branching fractions. The last sum, that describes uncertainties from lines recorded with detector Q, also includes the uncertainty $u(nf)$ in the normalisation factor $nf$ connecting different spectral regions. The intensity uncertainties from the statistical noise were determined using GFit. They varied between $0.001\%$ for the strong lines and $\sim20\%$ for the weak lines or self-absorbed lines. Most of the lines have uncertainties below $1\%$. When there was self-absorption, we corrected these lines and added the uncertainty from self-absorption to the intensity uncertainty. The calibration lamp uncertainty is $7\%$ and the uncertainty of the normalisation factor is $5\%$. From propagation of errors and using Eq. (\ref{Eq3}), the uncertainty of the transition probability or $f$-value is defined as, 
\begin{equation}\label{Eq5}
\Bigg(\frac{u(f_k)}{f_k} \Bigg)^2=\Bigg(\frac{u(A_k)}{A_k} \Bigg)^2= \Bigg(\frac{u(BF)}{BF} \Bigg)^2 + \Bigg(\frac{u(\tau)}{\tau} \Bigg )^2,
\end{equation}
where $u(\tau)$ is the uncertainty of the radiative lifetime of the upper level. In the cases where we used experimental lifetimes of \citet{Jonsson}, the uncertainties  vary between $5\%$ and $7\%$. For the theoretical lifetime uncertainties, we compared our theoretical values (to be described in the following section) with the experimental lifetimes available in the literature \citep{Kwiatkowski, Jonsson, Larsson1, Larsson2, Aldenius}. Figure \ref{Fig2} shows the comparison of the experimental lifetime values with the theoretical lifetime values that we calculated. The blue dashed line marks the $15\%$ and the black dashed line marks the $10\%$ difference. As the difference is small, we adopted $10\%$ relative uncertainty for the theoretical lifetimes.

\section{Theoretical method}
We performed our calculations using the multiconfiguration Hartree-Fock method (MCHF) \citep{Froese2, Froese3}. In this method , atomic state functions (ASF) $\Psi(\gamma LS)$ for the $LS$ terms are represented by linear combinations of configuration state functions (CSF);
\begin{equation}\label{Eq6}
\Psi (\gamma LS) = \sum\limits_i c_i \Phi(\gamma_iLS). 
\end{equation}
In the equation, $\gamma$ represents the electronic configurations and the quantum numbers other than $L$ and $S$. The configuration state functions $\Phi(\gamma_iLS)$ are built from one-electron orbitals and $c_i$ are the mixing coefficients. The mixing coefficients and the radial parts of the one-electron orbitals are determined by solving a set of equations that results from applying the variational principle to the energy expression associated with the ASFs. \\
\indent We started with a calculation of the ASFs describing terms of the configurations with $n$ up to nine and $l$ up to $g$ such as 3s$^2$, 3s3p, 3s3d, \ldots 3s9g. The calculation was done in the simplest approximation, where each ASF consists of only one CSF. All the ASFs were determined together in the same run and the calculation yielded a number of orbitals that were kept fixed in the proceeding calculations. \\
\indent Terms involving configurations with $n=8,9$ were not our prime target. However, we included these terms in the initial calculation to obtain orbitals that are spatially localised far away from the nucleus. This leads to a more complete and balanced orbital set. To improve the ASFs describing terms of the configurations with $n$ up to seven and $l$ up to $g$, such as 3s$^2$, 3s3p, 3s3d, \ldots 3s7g, we performed calculations with systematically enlarged CSF expansions. These expansions were formed from single and double replacements of orbitals in the reference configurations with orbitals in an active orbital set. We applied restrictions that there should be at most one replacement from 2s$^2$2p$^6$ and 1s$^2$ should be a closed shell. The orbitals in the active set were extended to include orbitals with $n=13$ and $l=h$. In these calculations, we determined ASFs with the same $LS$ symmetry together. \\
\indent Once the ASFs were determined, the oscillator strengths were calculated as expectation values of the transition operator. We performed the calculations both in the length and in the velocity gauge; see \citet{Froese2} for more details. For accurate calculations, the oscillator strengths in the two gauges should give the same value. In our calculations, the oscillator strengths in the two gauges typically agree to within $5\%$ for transitions between low-lying terms. The agreement is slightly worse for transitions involving the highest terms. Nevertheless, the velocity gauge, which weights more to the inner part of the wave function, shows good convergence properties and is believed to be the more accurate one for transitions involving the more excited states. \\
\indent All calculations were non-relativistic and the obtained $gf$ values represent term averages. To obtain the $gf$ values for the fine-structure transitions rather than for transitions between terms, we multiplied the $gf$ values for the term averages with the square of the line factor, see \citet{Cowan} Eq. (14.50).\\
\indent Moreover, we investigated the influence of relativistic effects by comparing our results with results from calculations where relativistic effects were accounted for in the Breit-Pauli approximation. As expected, the relativistic effects were insignificant with negligible term mixing. Exceptions are the $J=3$ states of the $^{1,3}$F terms in which the energy separations are so small that even weak relativistic effects give considerable term mixing. For the states of 4f $^{1,3}$F terms, we performed full calculations with relativistic effects in the Breit-Pauli method and applied the method of fine tuning  \citep{Brage} to match with the experimental data of \citet{Martin}. Due to the very small energy separations, it was not possible to perform these calculations for the higher $n$ and thus no theoretical oscillator strength values are given for states of the 5f, 6f, 7f $^{1,3}$F terms.

\section{Results and conclusions}

In this study, experimental and theoretical oscillator strengths of \ion{Mg}{I} are provided. $BF$s were obtained using Eq. \ref{Eq2} from the observed line intensities. We recorded the spectra using different currents and detectors. Applying different currents helped to rule out any self-absorption effects. The spectra, which are recorded with different detectors, are put on the same intensity scale by using a normalisation factor. In this way, we had all the lines from the same upper level on the same intensity scale. In the cases where we had unobservable weak lines, we used the theoretical transition probabilities to estimate the residual values.  \\
\indent From the measured $BF$s and radiative lifetimes, the transition probabilities, $A_{ul}$, are derived using Eq. \ref{Eq3} and $\log(gf)$ values are derived from Eq. \ref{Eq1}. For the experimental lifetimes, we used  the values of \citet{Jonsson}, and for others, we used our theoretical radiative lifetimes. Table 1 shows the theoretical lifetime values we computed together with the previous experimental and theoretical lifetime values. One notes that sometimes there are very large differences between values by \citet{Kurucz} and the experimental values.\\
\begin{figure}
\centering
\includegraphics[width=0.52\textwidth]{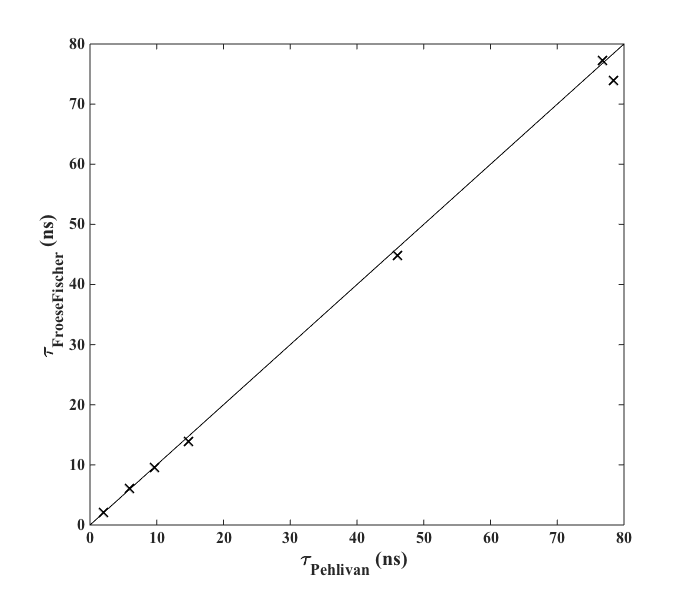}
\caption{Comparison between lifetime values of \citet{Froese} and  this work.}
\label{Fig4}
\end{figure}
\begin{figure}
\centering
\includegraphics[width=0.52\textwidth]{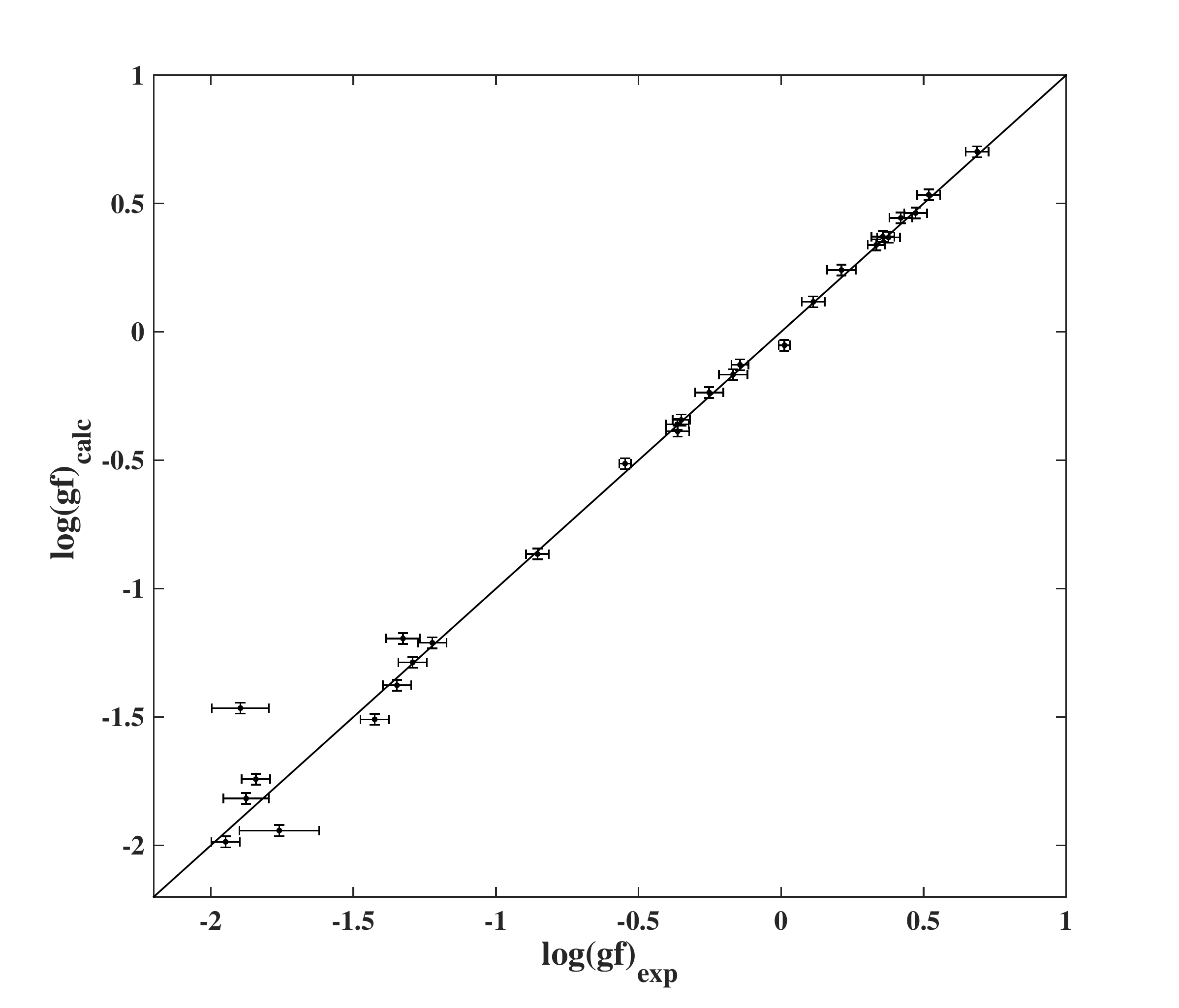}
\caption{Comparison between experimental and theoretical log$(gf)$ values of this work. The theoretical and experimental $\log(gf)$ values differ markedly for two transitions. These two transitions are affected by a blend and thus the derived experimental values are very uncertain (see text for more details).}
\label{Fig3}
\end{figure}
\begin{table*}
\caption{Theoretical radiative lifetimes of this work together with previous theoretical and experimental lifetimes. }
\label{table:1}      
\centering          
\begin{tabular}{l r r r | c c c c c}      
\hline\hline

Level & \multicolumn{3}{c}{Calculations (ns)} & \multicolumn{5}{c} {Experimental (ns)} \\
	 &    This work & K09$^1$    &CFF06$^2$      		      & KW80$^3$ & J84$^4$ & LS93$^5$ & L93$^6$ & A07$^7$ \\
	 \hline
	 \\
3s3p  $ {}^1$P$^o$  & 2. 1 & 1.7 & 2.1& -  & - & - & -  & -\\
3s4s  $ {}^3$S          & 9.6 & 7.6 & 9.6 & 9.7(6)  & - & - & -  & 11.5(1.0)\\
3s4s  $ {}^1$S          & 46 & 35 & 44.8& -  & 47(3) & - & -  & - \\
3s3d  $ {}^1$D          & 77 & 53 & 77.2&  -  &81(6) & - & -  & - \\
3s4p  $ {}^3$P$^o$   & 79 & 69 & 73.9& -  & - & - & -  & -\\
3s3d  $ {}^3$D          & 5.9 & 4.7 & 6.0&  5.9(4) & - & - & -  & 5.9(4) \\
3s4p  $ {}^1$P$^o$   & 14.7 & 16.8 & 13.8 &-  & - & 13.4(4) & -  & -\\
3s5s  $ {}^3$S          & 26 & 28 & -& -  & - & - & -  & 29(3)\\
3s5s  $ {}^1$S          & 102 & 65 &- & -  & 100(5) & - & -  & - \\
3s4d  $ {}^1$D          & 53 & 64 &- & -  & 57(3) & - & -  & - \\
3s4d  $ {}^3$D          & 16.1 & 13.7 &- & 15.6(9) & - & - & -  & 17.6(1.2) \\
3s5p  $ {}^3$P$^o$   & 268 & 211 &- & -  & - & - & -  & -\\
3s4f   $ {}^1$F$^o$   & 61 & 41 &- & -  & - & - & -  & -\\
3s4f   $ {}^3$F$^o$   & 61 & 48 &- & -  & - & - & -  & -\\
3s5p  $ {}^1$P$^o$   & 56 & 98 & - &-  & - & - & 54(3)  & -\\
3s6s  $ {}^3$S          & 57 & 63 &- & 51.8(3.0)  & - & - & -  & -\\
3s6s  $ {}^1$S          & 203 & 112 & - &-  & 211(12) & - & -  & -\\
3s5d  $ {}^1$D          & 43 & 56 & - &-  & 50(4) & - & -  & - \\
3s5d  $ {}^3$D          & 35 & 33 &- & 34.1(1.5)  & - & - & -  & 33(3) \\
3s6p  $ {}^3$P$^o$   & 642 & 502 &- & -  & - & - & -  & -\\
3s5f  $ {}^1$F$^o$   & 121 & 89 &- & -  & - & - & -  & -\\
3s5f  $ {}^3$F$^o$   & 119 & 102 &- & -  & - & - & -  & -\\
3s6p  $ {}^1$P$^o$   & 141 & 348 &- & -  & - & - & 140(10) & -\\
3s5g  $ {}^3$G          &226 & 211 & - &-  & - & - & -  & - \\
3s5g  $ {}^1$G          &226 & 211 &- & -  & - & - & -  & - \\
3s7s  $ {}^3$S          & 109 & 117 &- & -  & - & - & -  & -\\
3s7s  $ {}^1$S          & 366 & 181 &- & -  & 350(16)& - & -  & -\\
3s6d  $ {}^1$D          & 52 & 77 &- & -  & 54(3) & - & -  & - \\
3s6d  $ {}^3$D          & 65 & 68 & - &55.7(3.0)  & - & - & -  & - \\
3s7p  $ {}^3$P$^o$   & 1280 & 990 &- & -  & - & - & -  & -\\
3s6f  $ {}^1$F$^o$   & 216 & 165 &- & -  & - & - & -  & -\\
3s6f  $ {}^3$F$^o$   & 209 & 182 & - &-  & - & - & -  & -\\
3s7p  $ {}^1$P$^o$   & 291 & 752 & - &-  & - & - &290(20) & -\\
3s6g  $ {}^3$G          &387 & 365 &- & -  & - & - & -  & - \\
3s6g  $ {}^1$G          &387 & 365 &- & -  & - & - & -  & - \\
3s7d  $ {}^1$D          & 69 & 120 & - &-  & 70(6) & - & -  & - \\
3s7d  $ {}^3$D          & 113 & 126 &- &91.5(5.0) & - & - & -  & - \\
3s7f  $ {}^3$F$^o$   & 337 & 296 & - &-  & - & - & -  & -\\
3s7f  $ {}^1$F$^o$   & 355 & 287 & - &-  & - & - & -  & -\\
3s7g  $ {}^3$G          &610 & 578 & - &-  & - & - & -  & - \\
3s7g  $ {}^1$G          &610 & 578 & - &-  & - & - & -  & - \\
\\

 \hline

\end{tabular}
\tablebib{  $^{1}$\citet{Kurucz}; $^{2}$\citet{Froese}; $^{3}$\citet{Kwiatkowski}; $^{4}$\citet{Jonsson}; $^{5}$\citet{Larsson1}; $^{6}$\citet{Larsson2}; $^{7}$\citet{Aldenius}.}
\end{table*}
\indent Figure 2 shows a comparison of our theoretical lifetime values with experimental work of \citet{Kwiatkowski, Jonsson, Larsson1, Larsson2, Aldenius}. Overall, our calculations agree with the previously published experimental values within the 10\% uncertainty. Furthermore, we compared our lifetime values with the theoretical values from \citet{Froese} in Fig. \ref{Fig4} and the agreement is very good. Even the largest deviations are less than $6\%$.   \\
\indent From experimental $BF$s, we derived 34 $\log(gf)$ values of \ion{Mg}{I} lines from the upper even parity 4s $^{1,3}$S, 5s $^1$S, 3d $^1$D, and 4d $^1$D, and odd parity 4p $^3$P$^o$, 5p $^3$P$^o$, 4f $^{1,3}$F$^o$, and 5f $^{1,3}$F$^o$ with uncertainties in \textit{gf} as low as 5\%. In addition, we calculated theoretical $\log(gf)$ values of \ion{Mg}{I} lines up to $n=7$ from even parity $^{1,3}$S, $^{1,3}$D, and $^{1,3}$G terms, and odd parity $^{1,3}$P$^o$ and $^{1,3}$F$^o$ terms using ATSP2K package.  Figure \ref{Fig3} shows the comparison between the experimental and the theoretical $\log(gf)$ values. The good agreement between our experimental and theoretical $\log(gf)$ values makes us confident to recommend our theoretical values for the transitions in Table 3. Table 2 shows our experimental $\log(gf)$ values together with their uncertainty and corresponding theoretical $\log(gf)$ values that we calculated in this study, together with the branching fractions $BF$ and the transition probabilities $A_{ul}$. In addition, we compared our theoretical $\log(gf)$ values with \citet{Froese} values in Fig. \ref{Fig5}.  \citet{Froese} performed calculations for only the lowest lying levels up to $n=4$ while the current study calculations are additionally for higher levels up to $n=7$. The good agreement between our values and the theoretical values of \citet{Froese} is an additional indication of the quality of our values. Covering much more states and transitions, our calculations complement those of \citet{Froese}.  \\
\begin{figure}
\centering
\includegraphics[width=0.52\textwidth]{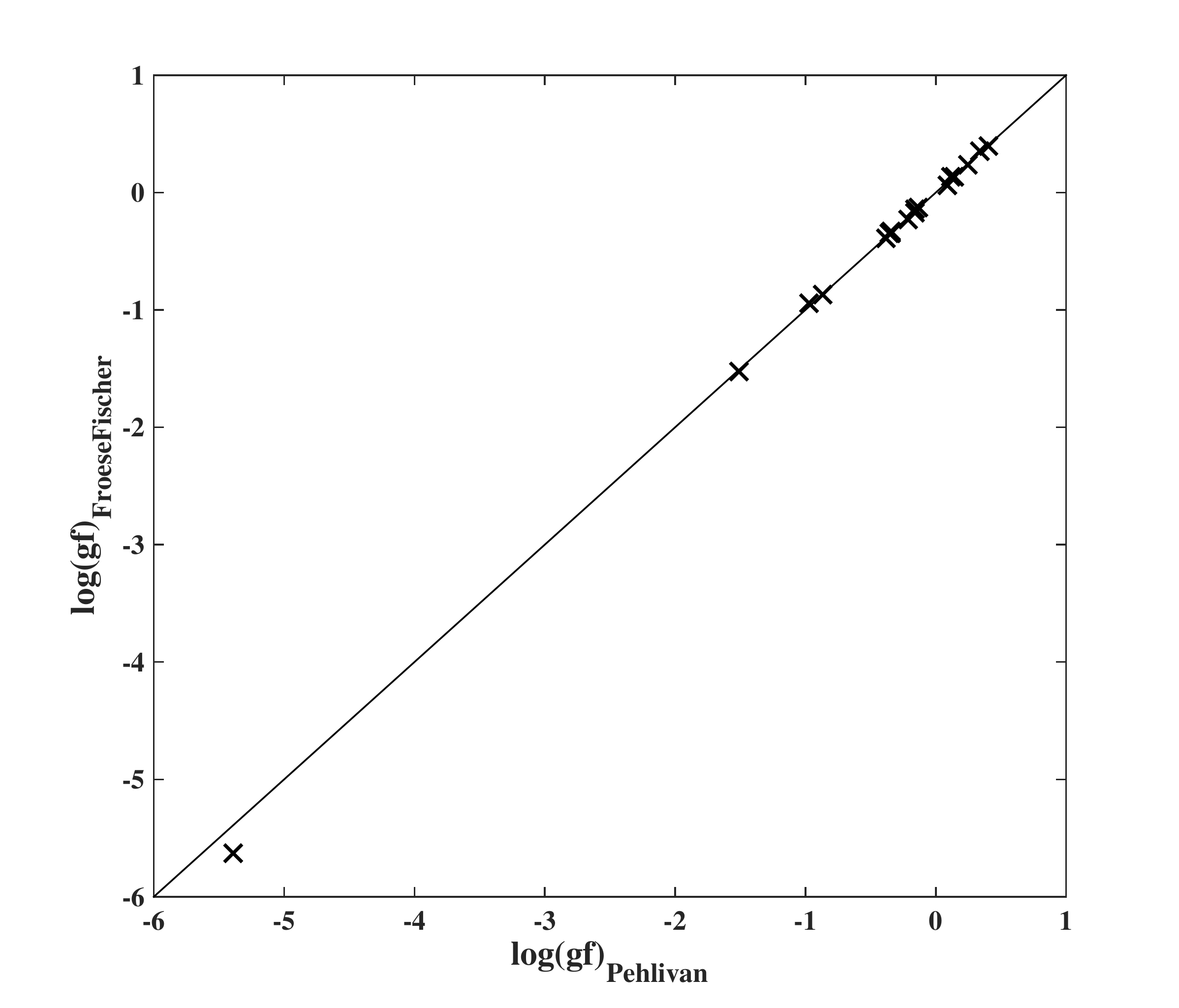}
\caption{Comparison of $\log(gf)$ values of the current study with the values of \citet{Froese}.}
\label{Fig5}
\end{figure}
\indent Overall, our theoretical lifetime values are in very good agreement with the experimental lifetime values in the literature. In addition, our theoretical $\log(gf)$ values agree with the experimental values of this work. However, our $\log(gf)$ values differ from the \citet{Aldenius} values for the optical \ion{Mg}{I} triplet lines (3p $^3$P$^o_{0,1,2}$ - 4s $^3$S$_1$), although we measured the same $BF$s. Figure \ref{Fig7} shows these lines in one of our spectra. The difference in $\log(gf)$ arises from the radiative lifetime of the upper level. \citet{Aldenius} measured the lifetime of $4\text{s}$ $^3$S$_1$ level to 11.5$\pm$1.0 ns.  
Other experimental studies find the lifetime ranging from 5.8 ns to 14.8 ns \citep{Berry, Schaefer, Andersen, Havey, Kwiatkowski}. Apparently, there is a large spread in the literature values and a 2 ns difference in lifetime corresponding to a 0.08 dex difference in log($gf$) values. The derived $\log(gf)$ value is thus sensitive to the choice of lifetime. Using the facility at Lund High Power Laser Centre, we remeasured the lifetime of this level. The atomic structure of \ion{Mg}{I} and technical limitations prevented us from deriving a conclusive value. However, the remeasured value leans towards the measurements by \citet{Kwiatkowski} and our calculated value. Therefore, we adopted our theoretical lifetime value (9.63 ns) for the $4\text{s}$ $^3$S$_1$ level. Our calculated lifetime value of 9.63 ns is a good choice, because it shows internal consistency between the length and velocity gauges, and in comparisons with other levels. Furthermore, \citet{Mashonkina} investigated the atomic data used in stellar magnesium abundance analyses. The paper found that the \citet{Aldenius} values overestimate the magnesium abundance by 0.11 dex compared to the other lines. With our experimental values this difference will be reduced.\\
\indent We recommend our experimental oscillator strengths when available. However, we would like to point out that the uncertainties of the \ion{Mg}{I} 15886.26 $\AA$ (6293.03 cm$^{-1}$) and 15886.18 $\AA$ (6293.06 cm$^{-1}$) are larger than 20\% owing to the weak line intensities and the blending of these two lines with each other. The oscillator strengths of these lines are outliers in Fig. \ref{Fig3}. The lines are displayed in figure 7. It is seen that they are weak with small separations, making fits to line profiles difficult. For these reasons, we advise the use of theoretical values for these transitions. When the experimental data are not available, we suggest theoretical values to be used.\\

\begin{figure*}
        \begin{subfigure}{0.50\textwidth}
                \includegraphics[width=\textwidth]{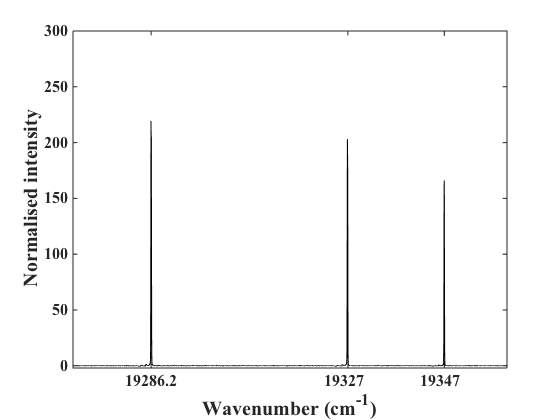}
                \caption{}
                \label{Fig7}
        \end{subfigure}
        \begin{subfigure}{0.50\textwidth}
                \includegraphics[width=\textwidth]{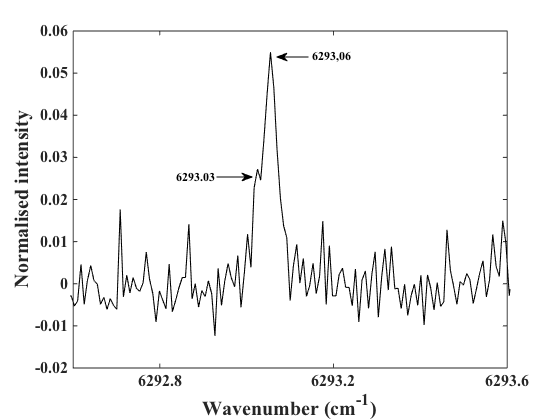}
                \caption{}
                \label{Fig6}
        \end{subfigure}
     \caption{Sample spectrum of two spectral regions of the FTS recordings. a) The \ion{Mg}{I} optical triplet lines (3p $^3$P$^o_{0,1,2}$ - 4s $^3$S$_1$). b) The feature consisting of the 3d $^3$D$_{2,1}$ - 5p $^3$P$^o_1$ transitions.}
\end{figure*}

\begin{longtab}
\begin{longtable}{l c c c c c c c }
\centering

Transition & $\lambda$ & $\sigma$       & $BF$ & $BF$ Unc. & $A_{ul}$ & log($gf$) & log($gf$)  \\ 
                &  $(\AA)$               &   (cm$^{-1}$)  &          & $\%$          & (s$^{-1}$)& Exp. & Calc. \\
\hline
\\
  3p $^3$P$^o_0$ - 4s $^3$S$_1$ &  5167.321  & 19346.997 & 0.1121 & 7 & $9.75 \times 10^6$  & -0.854$\pm$0.05 & -0.865\\
  3p $^3$P$^o_1$                           &   5172.684 & 19326.939 & 0.3471 & 3 & $3.02 \times 10^7$  & -0.363$\pm$0.04 & -0.387 \\    
  3p $^3$P$^o_2$ 			      &  5183.604  & 19286.225  & 0.5408 & 1 & $4.70 \times 10^7$  & -0.168$\pm$0.04 & -0.166 \\
  $Residual$				     & 		        	&		       & $<10^{-4}$&                                      &                             &       \\
  \multicolumn{1}{c} { $\tau=9.6^*$ ns  }          		     & 			 & 		      & 	    &    &            			   &				&    \\          
  \\
  3p $^1$P$^o_1$ - 4s $^1$S$_0$ & 11828.185 & 8452.08      & 1           & 0 & $2.13 \times10^7$  & -0.350$\pm$0.03 & -0.343 \\
  $Residual$				     & 		        	&		       & $<10^{-4}$&                                      &                             &       \\
    \multicolumn{1}{c} { $\tau=47$ ns  }          		     & 			 & 		      & 	    &    &            			   &				&    \\          
\\
  3p $^1$P$^o_1$ - 3d $^1$D$_2$ & 8806.756  & 11351.800   & 1           & 0 & $1.23 \times 10^7$  &  -0.144$\pm$0.03 & -0.128 \\
  $Residual$				     & 		        	&		       & $<10^{-4}$&                                      &                             &       \\
      \multicolumn{1}{c} { $\tau=81$ ns  }          		     & 			 & 		      & 	    &    &            			   &				&    \\          
\\
  4s $^3$S$_1$ - 4p $^3$P$^o_0$  & 15047.705 & 6643.71      & 1          & 0 & $1.27 \times 10^7$    & -0.364$\pm$0.04 & -0.360 \\
        \multicolumn{1}{c} { $\tau=78.5^*$ ns  }          		     & 			 & 		      & 	    &    &            			   &				&    \\         
\\
 4s $^3$S$_1$ - 4p $^3$P$^o_1$  & 15040.246 & 6647.01     & 1           &  0 & $1.27\times 10^7$     & 0.113$\pm$0.04 & 0.117  \\
  $Residual$				     & 		        	&		       & $<10^{-4}$&                                      &                             &       \\
        \multicolumn{1}{c} { $\tau=78.5^*$ ns  }          		     & 			 & 		      & 	    &    &            			   &				&    \\         
\\
 4s $^3$S$_1$ - 4p $^3$P$^o_2$  & 15024.992 & 6653.76      & 1          &  0  & $1.27\times 10^7$     & 0.334$\pm$0.03 & 0.339 \\
       \multicolumn{1}{c} { $\tau=78.5^*$ ns  }          		     & 			 & 		      & 	    &    &            			   &				&    \\         
 \\
 3p $^1$P$^o_1$ - 5s $^1$S$_0$ & 5711.088   & 17504.942  & 0.2942 & 10  & $2.94\times 10^6$    & -1.842$\pm$0.05   & -1.742 \\
 4p $^1$P$^o_1$                           & 31157.72$^{vac}$   & 3209.447    & 0.7057 & $ 1 $ & $7.06 \times 10^6$ & 0.012$\pm$0.02   & -0.052 \\
  $Residual$				     & 		        	&		       & $<10^{-4}$&                                      &                             &       \\
         \multicolumn{1}{c} { $\tau=100$ ns  }          		     & 			 & 		      & 	    &    &            			   &				&    \\         
  \\
 3p $^1$P$^o_1$ - 4d $^1$D$_2$ & 5528.405 & 18083.378  & 0.7060  & 3 & $1.24 \times 10^7$    & -0.547$\pm$0.02  & -0.513 \\
 4p $^1$P$^o_1$ 			     & 26399.76$^{vac}$ & 3787.88    & 0.2938  & 7 & $5.16 \times 10^6$    &  0.430$\pm$0.04  & 0.444 \\
  $Residual$				     & 		        	&		     & 0.0001  &                                          &                             &       \\
           \multicolumn{1}{c} { $\tau=57$ ns  }          		     & 			 & 		      & 	    &    &            			   &				&    \\    
\\
 4s $^3$S$_1$ - 5p $^3$P$^o_0$ & 7659.901  & 13051.405 & 0.3424  & 9 & $1.28 \times 10^6$ & -1.948$\pm$0.05 & -1.986 \\
 3d $^3$D$_1$ 			    & 15889.485 & 6291.74    & 0.0939  & 17 & $3.51 \times 10^5$ & -1.876$\pm$0.08 & -1.817 \\
 5s $^3$S$_1$				    & 42082.53$^{vac}$   & 2376.305  & 0.5636  & 5 & $2.11 \times 10^6$ & -0.252$\pm$0.05 & -0.236 \\
  $Residual$			            & 		        	&		    & $<10^{-4}$&                                  &                             &       \\
           \multicolumn{1}{c} { $\tau=267.6^*$ ns  }          		     & 			 & 		      & 	    &    &            			   &				&    \\    
\\
4s $^3$S$_1$ - 5p $^3$P$^o_1$ & 7659.152   & 13052.683  & 0.3810 & 9  & $1.42 \times 10^6$ & -1.425$\pm$0.05 & -1.509 \\
3d $^3$D$_2$ 				   & 15886.183  & 6293.06     & 0.0300 & 24 & $1.12 \times 10^5$ & -1.896$\pm$0.10 & -1.465 \\
3d $^3$D$_1$ 				   & 15886.261  & 6293.028   & 0.0410 & 38 & $1.53 \times 10^5$ &  -1.760$\pm$0.14 & -1.942 \\
5s $^3$S$_1$				   & 42059.93$^{vac}$    & 2377.595   & 0.5480 & 6   & $2.05 \times 10^6$ &   0.212$\pm$0.05  & 0.241 \\
  $Residual$				     & 		        	&		     & 0.0001  &                                          &                             &       \\
             \multicolumn{1}{c} { $\tau=267.6^*$ ns  }          		     & 			 & 		      & 	    &    &            			   &				&    \\   
\\
4s $^3$S$_1$ - 5p $^3$P$^o_2$ & 7657.603   & 13055.323  & 0.3105 & 8  & $1.16 \times 10^6$ & -1.292$\pm$0.05 & -1.287 \\
3d $^3$D$_3$ 				   & 15879.567  & 6295.68     & 0.0669 & 11 & $2.50 \times 10^5$ & -1.326$\pm$0.06 & -1.194 \\
5s $^3$S$_1$				   & 42013.28$^{vac}$ & 2380.236 & 0.5995 & 4 & $2.24 \times 10^6$ & 0.472$\pm$ 0.04 & 0.463 \\
  $Residual$				   & 		        	&		     & 0.0231 &                                          &                             &       \\
             \multicolumn{1}{c} { $\tau=267.6^*$ ns  }          		     & 			 & 		      & 	    &    &            			   &				&    \\   
\\
3d $^1$D$_2$ - 4f $^1$F$^o_3$ & 12083.662 & 8273.38  & 0.9505 & 1 & $1.55 \times 10^7$ & 0.377$\pm$0.04 & 0.368 \\
3d $^3$D$_2$ 				  &  14878.191 & 6719.42  & 0.0158 & 8 & $2.58 \times 10^5$ &  -1.223$\pm$0.05 & -1.211 \\
  $Residual$				   & 		        	&		     & 0.0337 &                                          &                             &       \\
               \multicolumn{1}{c} { $\tau=61.2^*$ ns  }          		     & 			 & 		      & 	    &    &            			   &				&    \\  
\\
3d $^3$D$_1$ - 4f $^3$F$^o_2$ & 14877.752  & 6719.60     & 0.8388 & 0.03 & $6.126 \times 10^8$ &  0.357$\pm$0.04 & 0.371 \\
  $Residual$				   & 		        	&		     & 0.1612 &                                          &                             &       \\
                 \multicolumn{1}{c} { $\tau=61.2^*$ ns  }          		     & 			 & 		      & 	    &    &            			   &				&    \\  
\\
3d $^1$D$_2$ - 4f $^3$F$^o_3$ & 12083.278  & 8273.64     & 0.0180 & 8 & $2.94 \times 10^5$ & -1.347$\pm$0.05 & -1.376 \\
3d $^3$D$_2$ 			   	   & 14877.608  & 6719.67     & 0.8699 & 1 & $1.42 \times 10^7$ & 0.518$\pm$0.04 & 0.534 \\
  $Residual$				   & 		        	&		     & 0.1121 &                                          &                             &       \\
                 \multicolumn{1}{c} { $\tau=61.2^*$ ns  }          		     & 			 & 		      & 	    &    &            			   &				&    \\  
\\
3d $^3$D$_3$ - 4f $^3$F$^o_4$ & 14877.529  & 6719.71     & 0.9987 & 3 & $1.63 \times 10^7$ & 0.688$\pm$0.04 & 0.702 \\
  $Residual$				   & 		        	&		     & 0.0013 &                                          &                             &       \\
                   \multicolumn{1}{c} { $\tau=61.2^*$ ns  }          		     & 			 & 		      & 	    &    &            			   &				&    \\  
\\
3d $^1$D$_2$ - 5f $^1$F$^o_3$ & 9255.778   	& 10801.098 & 0.8724 & 1 & $7.22 \times 10^6$ & -0.187$\pm$0.04 & -  \\
4d $^1$D$_2$ 				   & 24572.92$^{vac}$    & 4069.51 & 0.0776 & 10 & $6.42 \times 10^5$ & -0.391$\pm$0.06 & - \\
   $Residual$				   & 		        	&		     & 0.0501 &                                          &                             &       \\
                    \multicolumn{1}{c} { $\tau=120.7^*$ ns  }          		     & 			 & 		      & 	    &    &            			   &				&    \\  
\\
3d $^3$D$_1$ - 5f $^3$F$^o_2$ & 10811.158  & 9247.170   &  0.6499 & 2 & $5.45 \times 10^6$ & -0.321$\pm$0.04 & - \\
4d $^3$D$_1$				   & 33201.71$^{vac}$ & 3011.893 & 0.1891 & 8 & $1.58 \times 10^6$ & 0.117$\pm$0.05 & - \\
    $Residual$				   & 		        	&		     & 0.1610 &                                          &                             &       \\
                     \multicolumn{1}{c} { $\tau=119.4^*$ ns  }          		     & 			 & 		      & 	    &    &            			   &				&    \\  
\\
3d $^3$D$_3$ - 5f $^3$F$^o_4$ & 10811.053  & 9247.260   & 0.8524 & 1 &  $7.14 \times 10^6$ &  0.052$\pm$0.04 &  - \\
4d $^3$D$_3$				  &  33199.99$^{vac}$ & 3012.049 & 0.1465 & 7 & $1.23 \times 10^6$ & 0.261$\pm$0.05 & -   \\
    $Residual$				   & 		        	&		     & 0.0011 &                                          &                             &       \\
                         \multicolumn{1}{c} { $\tau=119.4^*$ ns  }          		     & 			 & 		      & 	    &    &            			   &				&    \\ 

 \hline                            
\end{longtable} 
\end{longtab}


\begin{longtab}
\begin{longtable}{l c c r r}
\caption{Presentation of theoretical $\log(gf)$ values of this work together with the transition, wavenumber, $\sigma$, wavelength, $\lambda_{air}$, and the transition probability. The wavelength and wavenumber values are taken from the compilation of \citet{Kaufman}. For $2000$ \AA $ < \lambda < 20000$ \AA, the wavelength is given in vacuum, otherwise in air. } \\
\hline \hline
Transition & $\sigma$ & $\lambda$  & \multicolumn{1}{c}{A$_{ul}$} &  \multicolumn{1}{c}{log(gf)}  \\
		& (cm$^{-1}$)      & (\AA)   		& \multicolumn{1}{c}{(s$^{-1} $)}			 &   \\
\hline
\\
\endfirsthead   
\caption{Continued.}\\
\hline\hline       
Transition & $\sigma$ & $\lambda $ &  \multicolumn{1}{c}{A$_{ul}$}  & \multicolumn{1}{c}{log(gf)}  \\
		& (cm$^{-1}$)      & (\AA)  		& \multicolumn{1}{c}{(s$^{-1} $)	}		 &   \\
\hline
\\
\endhead
\hline
\endfoot

3s$^2$ $^1$S$_0$ -	7p $^1$P$^o_1$ & 	58580.20 &	1707.06$^{vac}$ &	2.53 $\times$	10$^6$ &	-2.482 \\
3s$^2$ $^1$S$_0$ -	6p $^1$P$^o_1$ &	57215.00 &	1747.79$^{vac}$ &	5.42 $\times$	10$^6$ &	-2.131 \\
3s$^2$ $^1$S$_0$ -	5p $^1$P$^o_1$ &	54706.57 &	1827.94$^{vac}$ &	1.44 $\times$	10$^7$ &	-1.667 \\
3s$^2$ $^1$S$_0$ -	4p $^1$P$^o_1$ &	49346.73 &	2025.82 &	5.73 $\times$	10$^7$ &	-0.979 \\
3p $^3$P$^o_0$ -	7d $^3$D$_1$ &	37468.38 &	2668.12 &	7.04 $\times$	10$^5$ &	-1.951 \\
3p $^3$P$^o_1$ -	7d $^3$D$_1$ &	37448.33 &	2670.35 &	5.28 $\times$	10$^5$ &	-2.076 \\
3p $^3$P$^o_1$ -	7d $^3$D$_2$ &	37448.32 &	2669.55 &	1.58 $\times$	10$^6$ &	-1.599 \\
3p $^3$P$^o_2$ -	7d $^3$D$_1$ &	37407.62 &	2673.25 &	 3.52 $\times$	10$^4$ &	-3.253 \\
3p $^3$P$^o_2$ -	7d $^3$D$_2$ &	37407.60 &	2669.55 &	5.28 $\times$	10$^5$ &	-2.076 \\
3p $^3$P$^o_2$ -	7d $^3$D$_3$ &	37407.59 &	2672.46 &	2.96 $\times$	10$^6$ &	-1.328 \\
3p $^3$P$^o_0$ -	6d $^3$D$_1$ &	36592.48 &	2731.99 &	1.27 $\times$	10$^6$ &	-1.675 \\
3p $^3$P$^o_1$ -	6d $^3$D$_1$ &	36572.41 &	2733.49 &	9.51 $\times$	10$^5$ &	-1.800 \\
3p $^3$P$^o_1$ -	6d $^3$D$_2$ &	36572.40 &	2733.49 &	2.85 $\times$    10$^6$ &	-1.323 \\
3p $^3$P$^o_2$ -	6d $^3$D$_1$ &	36531.70 &	2736.54 &	6.34 $\times$	10$^4$ &	-2.976 \\
3p $^3$P$^o_2$ -	6d $^3$D$_2$ &	36531.68 &	2736.54 &	9.51 $\times$	10$^5$ &	-1.800 \\
3p $^3$P$^o_2$ -	6d $^3$D$_3$ &	36531.66 &	2736.54 &	5.33 $\times$	10$^6$ &	-1.052 \\

3p $^3$P$^o_0$ -	7s $^3$S$_1$ &  	36004.81 &	2777.41 &	7.17 $\times$	10$^5$ &	-2.608 \\
3p $^3$P$^o_1$ -	7s $^3$S$_1$ &  	35984.75 &	2778.95 &	2.15 $\times$	10$^6$ &	-2.131 \\
3p $^3$P$^o_2$ -	7s $^3$S$_1$ &  	35943.96 &	2781.28 &	3.58 $\times$	10$^6$ &	-1.909 \\
3p $^3$P$^o_0$ -	5d $^3$D$_1$ &	35117.87 &	2846.72 &	2.50 $\times$	10$^6$ &	-1.344 \\
3p $^3$P$^o_1$ -	5d $^3$D$_2$ &	35097.83 &	2848.35 &	5.63 $\times$	10$^6$ &	-0.992 \\
3p $^3$P$^o_1$ -	5d $^3$D$_1$ &	35097.81 &	2848.34 &	1.88 $\times$	10$^6$ &	-1.469 \\
3p $^3$P$^o_2$ -	5d $^3$D$_1$ &	35057.09 &	2851.65 &	1.25 $\times$	10$^5$ &	-2.645 \\
3p $^3$P$^o_2$ -	5d $^3$D$_2$ &	35057.07 &	2851.65 &	1.88 $\times$	10$^6$ &	-1.469 \\
3p $^3$P$^o_2$ -	5d $^3$D$_3$ &	35056.99 &	2851.66 &	1.05 $\times$	10$^7$ &	-0.721 \\
3s$^2$ $^1$S$_0$ -	3p $^1$P$^o_1$ &	35051.25 &	2852.13 &	4.79 $\times$	10$^8$ &	0.240 \\
3p $^3$P$^o_0$ -	6s $^3$S$_1$ &  	34041.42 &	2936.74 &	1.40 $\times$	10$^6$ &	-2.269 \\
3p $^3$P$^o_1$ -	6s $^3$S$_1$ &  	34021.33 &	2938.47 &	4.20 $\times$	10$^6$ &	-1.792 \\
3p $^3$P$^o_2$ -	6s $^3$S$_1$ &  	33980.60 &	2941.99 &7.00 $\times$	10$^6$ &	-1.570 \\

3p $^3$P$^o_0$ -	4d $^3$D$_1$ &	32341.93 &	3091.06 &	5.82 $\times$	10$^6$ &	-0.907 \\
3p $^3$P$^o_1$ -	4d $^3$D$_1$ &	32321.87 &	3092.98 &	4.37 $\times$	10$^6$ &	-1.032 \\
3p $^3$P$^o_1$ -	4d $^3$D$_2$ & 	32321.85 &	3092.99 &	1.31 $\times$	10$^7$ &	-0.555 \\
3p $^3$P$^o_2$ -	4d $^3$D$_1$ &	32281.16 &	3096.88 &	2.91 $\times$	10$^5$ &	-2.208 \\
3p $^3$P$^o_2$ -	4d $^3$D$_2$ & 	32281.12 &	3096.89 &	4.37 $\times$	10$^6$ &	-1.032 \\
3p $^3$P$^o_2$ -	4d $^3$D$_3$ &	32281.09 &	3096.89 &	2.45 $\times$	10$^7$ &	-0.284 \\
3p $^3$P$^o_0$ -	5s $^3$S$_1$ & 	30022.13 &	3329.92 &	3.22 $\times$	10$^6$ &	-1.798 \\
3p $^3$P$^o_1$ -	5s $^3$S$_1$ & 	30002.06 &	3332.15 &	9.67 $\times$	10$^6$ &	-1.321 \\
3p $^3$P$^o_2$ -	5s $^3$S$_1$ & 	29961.35 &	3336.67 &	1.61 $\times$	10$^7$ &	-1.099 \\
3p $^3$P$^o_0$ -	3d $^3$D$_1$ &	26106.65 &	3829.35 &	1.88 $\times$	10$^7$ &	-0.214 \\
3p $^3$P$^o_1$ -	3d $^3$D$_1$ &	26086.59 &	3832.30 &	1.41 $\times$	10$^7$ &	-0.339 \\
3p $^3$P$^o_1$ -	3d $^3$D$_2$ & 	26086.56 &	3832.30 &	4.23 $\times$	10$^7$ &	0.138 \\
3p $^3$P$^o_2$ -	3d $^3$D$_1$ &	26045.88 &	3838.29 &	9.40 $\times$	10$^5$ &	-1.515 \\
3p $^3$P$^o_2$ -	3d $^3$D$_3$ &	26045.87 &	3838.29 &	7.89 $\times$	10$^7$ &	0.409 \\
3p $^3$P$^o_2$ -	3d $^3$D$_2$ & 	26045.85 &	3838.30 &	1.41 $\times$	10$^7$ &	-0.339 \\
3p $^1$P$^o_1$ - 	7d $^1$D$_2$ &	23989.76 &	4167.27 &	1.39 $\times$	10$^7$ &	-0.746 \\

3p $^1$P$^o_1$ - 	6d $^1$D$_2$ &	22971.98 &	4351.91 &	1.83 $\times$	10$^7$ &	-0.588 \\
3p $^1$P$^o_1$ -  	7s $^1$S$_0$ &  	22958.14 &	4354.53 &	5.34 $\times$	10$^5$ &	-2.820 \\
3s$^2$ $^1$S$_0$ -	3p $^3$P$_1$ &	21870.46 &	4571.10 &	3.94 $\times$ 	10$^2$ &	-5.397 \\
3p $^1$P$^o_1$ - 	5d $^1$D$_2$ & 	21257.12 &	4702.99 &	2.12 $\times$	10$^7$ &	-0.456 \\
3p $^1$P$^o_1$ -  	6s $^1$S$_0$ &  	21135.61 &	4730.03 &	1.25 $\times$	10$^6$ &	-2.379 \\
3p $^3$P$^o_0$ -	4s $^3$S$_1$ & 	19347.00 &	5167.32 &	1.15 $\times$	10$^7$ &	-0.865 \\
3p $^3$P$^o_1$ -	4s $^3$S$_1$ & 	19326.94 &	5172.68 &	3.46 $\times$	10$^7$ &	-0.387 \\
3p $^3$P$^o_2$ -	4s $^3$S$_1$ & 	19286.23 &	5183.60 &	5.77 $\times$	10$^7$ &	-0.166 \\
3p $^1$P$^o_1$ -  	4d $^1$D$_2$ & 	18083.38 &	5528.40 &	1.35 $\times$	10$^7$ &	-0.513 \\
3p $^1$P$^o_1$ - 	5s $^1$S$_0$ &  	17504.94 &	5711.09 &	3.71 $\times$	10$^6$ &	-1.742 \\
4s $^3$S$_1$ -	7p $^3$P$^o_2$ &	17280.36 &	5785.31 &	6.90 $\times$	10$^4$ &	-2.506 \\
4s $^3$S$_1$ -	7p $^3$P$^o_1$ &	17279.62 &	5785.56 &	4.14 $\times$	10$^4$ &	-2.728 \\
4s $^3$S$_1$ -	7p $^3$P$^o_0$ &	17279.29 &	5787.28 &	1.38 $\times$	10$^4$ &	-3.205 \\
4s $^3$S$_1$ -	6p $^3$P$^o_2$ &	15821.63 &	6318.72 &	1.77 $\times$	10$^5$ &	-2.020 \\
4s $^3$S$_1$ -	6p $^3$P$^o_1$ &	15820.32 &	6319.24 &	1.06 $\times$     10$^5$ &	-2.242 \\
4s $^3$S$_1$ -	6p $^3$P$^o_0$ &	15819.68 &	6319.50 &	3.54 $\times$	10$^4$ &	-2.719 \\
4s $^1$S$_0$ - 	7p $^1$P$^o_1$ & 	15076.90 &	6630.83 &	6.07 $\times$	10$^3$ &	-3.923 \\
4s $^1$S$_0$ - 	6p $^1$P$^o_1$ &	13711.65 &	7291.06 &	6.32 $\times$	10$^4$ &	-2.823 \\
4s $^3$S$_1$ -	5p $^3$P$^o_2$ &	13055.32 &	7657.60 &	6.51 $\times$	10$^5$ &	-1.287 \\
4s $^3$S$_1$ -	5p $^3$P$^o_1$ &	13052.68 &	7659.15 &	3.91 $\times$	10$^5$ &	-1.509 \\
4s $^3$S$_1$ -	5p $^3$P$^o_0$ & 	13051.41 &	7659.90 &	1.30 $\times$	10$^5$ &	-1.986 \\

4p $^3$P$^o_0$ - 	6s $^3$S$_1$ &  	12417.91 &	8050.66 &	3.17 $\times$	10$^5$ &	-1.662 \\
3d $^1$D$_2$   - 	7p $^1$P$^o_1$ & 	12177.16 &	8209.84 &	1.78 $\times$	10$^5$ &	-2.264 \\
4p $^3$P$^o_0$ -	7d $^3$D$_1$ &	11477.67 &	8710.17 &	1.39 $\times$	10$^5$ &	-1.629 \\
4p $^3$P$^o_1$ -	7d $^3$D$_1$ &	11474.38 &	8712.68 &	1.04 $\times$	10$^5$ &	-1.754 \\
4p $^3$P$^o_1$ -	7d $^3$D$_2$ &	11474.36 &	8712.69 &	3.12 $\times$    10$^5$ &	-1.277 \\
4p $^3$P$^o_2$ -	7d $^3$D$_1$ &	11467.63 &	8717.80 &	6.94 $\times$	10$^3$ &	-2.930 \\
4p $^3$P$^o_2$ -	7d $^3$D$_2$ &	11467.61 &	8717.82 &	1.04 $\times$	10$^5$ &	-1.754 \\
4p $^3$P$^o_2$ -	7d $^3$D$_3$ &	11467.60 &	8717.83 &	5.83 $\times$	10$^5$ &	-1.006 \\
3p $^1$P$^o_1$ - 	3d $^1$D$_2$ & 	11351.80 &	8806.76 &	1.30 $\times$	10$^7$ &	-0.128 \\
4s $^1$S$_0$ - 	5p $^1$P$^o_1$ &	11203.20 &	8923.57 &	5.87 $\times$	10$^5$ &	-1.679 \\
3d $^1$D$_2$   -	6p $^1$P$^o_1$ &	10811.94 &	9246.51 &	2.72 $\times$	10$^5$ &	-1.976 \\
4p $^3$P$^o_0$ -	6d $^3$D$_1$ &	10601.75 &	9429.81 &	2.49 $\times$	10$^5$ &	-1.306 \\
4p $^3$P$^o_1$ -	6d $^3$D$_1$ &	10598.46 &	9432.75 &	1.87 $\times$	10$^5$ &	-1.431 \\
4p $^3$P$^o_1$ -	6d $^3$D$_2$ &	10598.44 &	9432.76 &	5.61 $\times$	10$^5$ &	-0.954 \\
4p $^3$P$^o_2$ -	6d $^3$D$_1$ &	10591.71 &	9438.76 &	1.25 $\times$	10$^4$ &	-2.607 \\
4p $^3$P$^o_2$ -	6d $^3$D$_2$ &	10591.69 &	9438.77 &	1.87 $\times$	10$^5$ &	-1.431 \\
4p $^3$P$^o_2$ -	6d $^3$D$_3$ &	10591.68 &	9438.78 &	1.05 $\times$	10$^6$ &	-0.683 \\

3d $^3$D$_2$   -	7p $^3$P$^o_2$ &	10520.73 &	9505.04 &	5.69 $\times$	10$^3$ &	-3.154 \\
3d $^3$D$_3$   -	7p $^3$P$^o_2$ &	10520.71 &	9502.45 &	3.19 $\times$	10$^4$ &	-2.406 \\
3d $^3$D$_1$   -	7p $^3$P$^o_2$ &	10520.70 &	9505.07 &	3.79 $\times$	10$^2$ &	-4.330 \\
3d $^3$D$_2$   -	7p $^3$P$^o_1$ &	10519.99 &	9503.10 &	1.71 $\times$	10$^4$ &	-2.677 \\
3d $^3$D$_1$   -	7p $^3$P$^o_1$ &	10519.96 &	9505.74 &	5.69 $\times$	10$^3$ &	-3.154 \\
3d $^3$D$_1$   -	7p $^3$P$^o_0$ &	10519.63 &	9503.43 &	7.58 $\times$	10$^3$ &	-3.029 \\
4p $^3$P$^o_0$ -	7s $^3$S$_1$ &  	10014.08 &	9983.19 &	1.50 $\times$	10$^5$ &	-2.177 \\
4p $^3$P$^o_1$ -	7s $^3$S$_1$ &  	10010.80 &	9986.47 &	4.49 $\times$	10$^5$ &	-1.700 \\
4p $^3$P$^o_2$ -	7s $^3$S$_1$ &  	10004.05 &	9993.21 &	7.48 $\times$	10$^5$ &	-1.478 \\
4p $^1$P$^o_1$ -   	7d $^1$D$_2$ &	9694.28 &	10312.52 &	2.44 $\times$	10$^5$ &	-1.718 \\
4p $^3$P$^o_0$ -	5d $^3$D$_1$ &	9127.15 &	10953.32 &	4.90 $\times$	10$^5$ &	-0.883 \\
4p $^3$P$^o_1$ -	5d $^3$D$_1$ &	9123.86 &	10957.28 &	3.67 $\times$	10$^5$ &	-1.008 \\
4p $^3$P$^o_1$ -	5d $^3$D$_2$ &	9123.83 &	10957.30 &	1.10 $\times$  10$^6$ &	-0.531 \\
4p $^3$P$^o_2$ -	5d $^3$D$_1$ &	9117.11 &	10965.39 &	2.45 $\times$	10$^4$ &	-2.184 \\
4p $^3$P$^o_2$ -	5d $^3$D$_2$ &	9117.09 &	10965.41 &	3.67 $\times$	10$^5$ &	-1.008 \\
4p $^3$P$^o_2$ -	5d $^3$D$_3$ &	9117.06 &	10965.45 &	2.06 $\times$	10$^6$ &	-0.260 \\

3d $^3$D$_2$   -	6p $^3$P$^o_2$ &	9062.00 &	11032.07 &	1.24 $\times$	10$^4$ &	-2.686 \\
3d $^3$D$_3$   -	6p $^3$P$^o_2$ &	9061.97 &	11032.10 &	6.93 $\times$	10$^4$ &	-1.938 \\
3d $^3$D$_1$   -	6p $^3$P$^o_2$ &	9061.97 &	11032.11 &	8.25 $\times$	10$^2$ &	-3.862 \\
3d $^3$D$_2$   -	6p $^3$P$^o_1$ &	9060.69 &	11033.66 &	3.71 $\times$	10$^4$ &	-2.209 \\
3d $^3$D$_1$   -	6p $^3$P$^o_1$ &	9060.67 &	11033.69 &	1.24 $\times$	10$^4$ &	-2.686 \\
3d $^3$D$_1$   -	6p $^3$P$^o_0$ &	9060.00 &	11034.48 &	1.65 $\times$	10$^4$ &	-2.561 \\
4p $^1$P$^o_1$ -  	6d $^1$D$_2$ &	8676.49 &	11522.21 &	1.25 $\times$	10$^5$ &	-1.913 \\
4p $^1$P$^o_1$ -  	7s $^1$S$_0$ &  	8662.64 &	11540.61 &	9.02 $\times$	10$^5$ &	-1.745 \\
3p $^1$P$^o_1$ - 	4s $^1$S$_0$ &  	8452.08 &	11828.19 &	2.17 $\times$	10$^5$ &	-0.343 \\
3d $^1$D$_2$   - 	5p $^1$P$^o_1$ &	8303.50 &	12039.86 &	4.25 $\times$	10$^7$ &	-1.551 \\
3d $^1$D$_2$   -	4f $^3$F$^o_3$ &	8273.64 &	12083.28 &	2.95 $\times$	10$^5$ &	-1.376 \\
3d $^1$D$_2$   -	4f $^1$F$^o_3$ &	8273.38 &	12083.66 &	1.55 $\times$	10$^7$ &	0.368 \\
4p $^3$P$^o_1$ -	6s $^3$S$_1$ &  	8047.36 &	12422.99 &	9.50 $\times$	10$^5$ &	-1.185 \\
4p $^3$P$^o_2$ -	6s $^3$S$_1$ &  	8040.62 &	12433.42 &	1.58 $\times$	10$^6$ &	-0.964 \\
4p $^1$P$^o_1$ -   	5d $^1$D$_2$ & 	6961.63 &	14360.48 &	3.65 $\times$	10$^4$ &	-2.255 \\

4p $^1$P$^o_1$ -   	6s $^1$S$_0$ &  	6840.13 &	14615.58 &	1.84 $\times$	10$^6$ &	-1.230 \\
3d $^3$D$_3$   -	4f $^3$F$^o_4$ &	6719.71 &	14877.53 &	6.98 $\times$	10$^6$ &	0.702 \\
3d $^3$D$_2$   -	4f $^3$F$^o_3$ &	6719.67 &	14877.61 &	4.83 $\times$	10$^6$ &	0.534 \\
3d $^3$D$_3$   -	4f $^3$F$^o_3$ &	6719.66 &	14877.65 &	6.03 $\times$	10$^5$ &	-0.370 \\
3d $^3$D$_2$   -	4f $^3$F$^o_2$ &	6719.63 &	14877.71 &	6.03 $\times$	10$^5$ &	-0.362 \\
3d $^3$D$_3$   -	4f $^3$F$^o_2$ &	6719.61 &	14877.75 &	1.72 $\times$	10$^4$ &	-1.906 \\
3d $^3$D$_1$   - 	4f $^3$F$^o_2$ &	6719.60 &	14877.78 &	3.26 $\times$	10$^6$ &	0.371 \\
3d $^3$D$_2$   -	4f $^1$F$^o_3$ &	6719.41 &	14882.26 &	2.67 $\times$	10$^5$ &	-1.211 \\
3d $^3$D$_3$   -	4f $^1$F$^o_3$ &	6719.39 &	14882.30 &	3.35 $\times$	10$^4$ &	-2.113 \\
4s $^3$S$_1$ -	4p $^3$P$^o_2$ &	6653.76 &	15029.10 &	7.08 $\times$	10$^6$ &	0.339 \\
4s $^3$S$_1$ -	4p $^3$P$^o_1$ &	6647.01 &	15040.25 &	4.25 $\times$	10$^6$ &	0.117 \\
4s $^3$S$_1$ -	4p $^3$P$^o_0$ & 	6643.71 &	15047.71 &	1.42 $\times$	10$^6$ &	-0.360 \\
5s $^3$S$_1$ -	7p $^3$P$^o_2$ &	6605.26 &	15135.37 &	6.36 $\times$	10$^4$ &	-1.706 \\
5s $^3$S$_1$ -	7p $^3$P$^o_1$ &	6604.51 &	15137.07 &	3.82 $\times$	10$^4$ &	-1.928 \\
5s $^3$S$_1$ -	7p $^3$P$^o_0$ &	6604.15 &	15137.83 &	1.27 $\times$	10$^4$ &	-2.405 \\
4p $^3$P$^o_0$ -	4d $^3$D$_1$ &	6351.22 &	15740.72 &	1.09 $\times$	10$^6$ &	-0.223 \\
4p $^3$P$^o_1$ -	4d $^3$D$_1$ &	6347.92 &	15748.89 &	8.19 $\times$	10$^5$ &	-0.348 \\
4p $^3$P$^o_1$ -	4d $^3$D$_2$ & 	6347.88 &	15748.99 &	2.46 $\times$	10$^6$ &	0.129 \\
4p $^3$P$^o_2$ -	4d $^3$D$_1$ &	6341.17 &	15765.65 &	5.46 $\times$	10$^4$ &	-1.524 \\
4p $^3$P$^o_2$ -	4d $^3$D$_2$ & 	6341.13 &	15765.75 &	8.19 $\times$	10$^5$ &	-0.348 \\
4p $^3$P$^o_2$ -	4d $^3$D$_3$ &	6341.10 &	15765.84 &	4.59 $\times$	10$^6$ &	0.400 \\

3d $^3$D$_2$   -	5p $^3$P$^o_2$ &	6295.70 &	15879.52 &	3.28 $\times$	10$^4$ &	-1.942 \\
3d $^3$D$_3$   -	5p $^3$P$^o_2$ &	6295.68 &	15879.57 &	1.84 $\times$	10$^5$ &	-1.194 \\
3d $^3$D$_1$   -	5p $^3$P$^o_2$ &	6295.67 &	15879.60 &	2.19 $\times$	10$^3$ &	-3.118 \\
3d $^3$D$_2$   -	5p $^3$P$^o_1$ &	6293.06 &	15886.18 &	9.84 $\times$	10$^4$ &	-1.465 \\
3d $^3$D$_1$   -	5p $^3$P$^o_1$ &	6293.03 &	15886.26 &	3.28 $\times$	10$^4$ &	-1.942 \\
3d $^3$D$_1$   -	5p $^3$P$^o_0$ & 	6291.74 &	15889.49 &	4.38 $\times$	10$^4$ &	-1.817 \\
5s $^1$S$_0$ - 	7p $^1$P$^o_1$ & 	6024.05 &	16595.68 &	6.62 $\times$	10$^4$ &	-2.090 \\
4s $^1$S$_0$ - 	4p $^1$P$^o_1$ &	5843.41 &	17108.66 &	9.37 $\times$  10$^6$ &	0.090 \\
4d $^1$D$_2$ - 	7p $^1$P$^o_1$ & 	5445.62 &	18358.50 &	1.17 $\times$	10$^5$ &	-1.750 \\
5s $^3$S$_1$ -	6p $^3$P$^o_2$ &	5146.53 &	19425.38 &	1.84 $\times$	10$^5$ &	-1.027 \\
5s $^3$S$_1$ -	6p $^3$P$^o_1$ &	5145.21 &	19430.29 &	1.11 $\times$	10$^5$ &	-1.249 \\
5s $^3$S$_1$ -	6p $^3$P$^o_0$ &	5144.57 &	19432.73 &	3.68 $\times$	10$^4$ &	-1.726 \\
5p $^3$P$^o_0$ -	7d $^3$D$_1$ &	5070.00 &	19718.54 &	6.28 $\times$	10$^4$ &	-1.263 \\
5p $^3$P$^o_1$ -	7d $^3$D$_1$ &	5068.71 &	19723.51 &	4.71 $\times$	10$^4$ &	-1.388 \\
5p $^3$P$^o_1$ -	7d $^3$D$_2$ &	5068.70 &	19723.58 &	1.41 $\times$	10$^5$ &	-0.911 \\
5p $^3$P$^o_2$ -	7d $^3$D$_1$ &	5066.07 &	19733.79 &	3.14 $\times$	10$^3$ &	-2.564 \\
5p $^3$P$^o_2$ -	7d $^3$D$_3$ &	5066.05 &	19733.90 &	2.64 $\times$	10$^5$ &	-0.640 \\
5p $^3$P$^o_2$ -	7d $^3$D$_2$ &	5066.05 &	19733.86 &	4.71 $\times$	10$^4$ &	-1.388 \\
4f $^3$F$^o_3$ -	7g $^3$G$_4$ &	4747.10 &	21065.50$^{vac}$&	5.38 $\times$	10$^5$ &	-0.494 \\
4f $^1$F$^o_3$ - 	7g $^3$G$_3$ &	4747.10 &	21065.50$^{vac}$ &	1.04 $\times$	10$^3$ &	-3.316 \\
4f $^1$F$^o_3$ - 	7g $^3$G$_4$ &	4747.10 &	21065.50$^{vac}$ &	1.47 $\times$	10$^5$ &	-1.057 \\

4f $^3$F$^o_2$ -	7g $^3$G$_3$ &	4746.90 &	21065.50$^{vac}$ &	6.67 $\times$	10$^5$ &	-0.510 \\
4f $^1$F$^o_3$ - 	7g $^1$G$_4$ &	4746.84 &	21066.66$^{vac}$ &	5.78 $\times$	10$^5$ &	-0.463 \\
4f $^3$F$^o_3$ -	7g $^1$G$_4$ &	4746.84 &	21066.66$^{vac}$ &	1.43 $\times$	10$^5$ &	-1.069 \\
4f $^3$F$^o_3$ -	7g $^3$G$_3$ &	4746.84 &	21066.66$^{vac}$ &	5.73 $\times$	10$^4$ &	-1.576 \\
4f $^3$F$^o_4$ -	7g $^3$G$_5$ &	4746.80 &	21066.90$^{vac}$ &	7.26 $\times$	10$^5$ &	-0.277 \\
4f $^3$F$^o_4$ -	7g $^3$G$_3$ &	4746.78 &	21066.90$^{vac}$ &	9.25 $\times$	10$^2$ &	-3.368 \\
4f $^3$F$^o_4$ -	7g $^3$G$_4$ &	4746.78 &	21066.90$^{vac}$ &	4.04 $\times$	10$^4$ &	-1.619 \\
4f $^3$F$^o_4$ - 	7g $^1$G$_4$ &	4746.78 &	21066.90$^{vac}$ &	4.99 $\times$	10$^4$ &	-2.527 \\
5s $^1$S$_0$ - 	6p $^1$P$^o_1$ &	4658.81 &	21464.82$^{vac}$&	2.33 $\times$	10$^5$ &	-1.322 \\
4f $^1$F$^o_3$ - 	7d $^3$D$_3$ &	4642.33 &	21540.93$^{vac}$ &	4.82 $\times$	10$^1$ &	-4.632 \\
4f $^3$F$^o_2$ -	7d $^3$D$_1$ &	4642.14 &	21541.79$^{vac}$ &	8.83 $\times$	10$^3$ &	-2.040 \\
4f $^3$F$^o_2$ -	7d $^3$D$_2$ &	4642.12 &	21541.88$^{vac}$ &	1.64 $\times$	10$^3$ &	-2.773 \\
4f $^3$F$^o_2$ -	7d $^3$D$_3$ &	4642.11 &	21541.93$^{vac}$ &	7.74 $\times$	10$^1$ &	-4.427 \\
4f $^3$F$^o_3$ -	7d $^3$D$_2$ &	4642.07 &	21542.10$^{vac}$ &	1.31 $\times$	10$^4$ &	-1.869 \\
4f $^3$F$^o_3$ -	7d $^3$D$_3$ &	4642.06 &	21542.15$^{vac}$ &	2.66 $\times$	10$^3$ &	-2.890 \\
4f $^3$F$^o_4$ -	7d $^3$D$_3$ &	4642.01 &	21542.40$^{vac}$ &	3.14 $\times$	10$^4$ &	-1.819 \\
4f $^1$F$^o_3$ - 	7d $^1$D$_2$ &	4364.58 &	22911.71$^{vac}$ &	7.07 $\times$	10$^4$ &	-1.568 \\
5p $^1$P$^o_1$ -  	7d $^1$D$_2$ &	4334.48 &	23070.80$^{vac}$&	3.92 $\times$	10$^3$ &	-3.817 \\
4d $^3$D$_3$ -	7p $^3$P$^o_2$ &	4285.51 &	23334.48$^{vac}$ &	3.52 $\times$	10$^4$ &	-1.583 \\
4d $^3$D$_2$ -	7p $^3$P$^o_2$ &	4285.47 &	23334.69$^{vac}$ &	6.29 $\times$	10$^3$ &	-2.331 \\
4d $^3$D$_1$ -	7p $^3$P$^o_2$ &	4285.43 &	23334.91$^{vac}$ &	4.19 $\times$	10$^2$ &	-3.507 \\

4d $^3$D$_2$ -	7p $^3$P$^o_1$ &	4284.73 &	23338.72$^{vac}$ &	1.89 $\times$	10$^4$ &	-1.854 \\
4d $^3$D$_1$ -	7p $^3$P$^o_1$ &	4284.69 &	23338.94$^{vac}$ &	6.29 $\times$	10$^3$ &	-2.331 \\
4d $^3$D$_1$ - 	7p $^3$P$^o_0$ &	4284.35 &	23340.74$^{vac}$ &	8.38 $\times$	10$^3$ &	-2.206 \\
5p $^3$P$^o_0$ -	6d $^3$D$_1$ &	4194.07 &	23843.22$^{vac}$ &	1.11 $\times$	10$^5$ &	-0.851 \\
5p $^3$P$^o_1$ -	6d $^3$D$_1$ &	4192.79 &	23850.48$^{vac}$ &	8.34 $\times$	10$^4$ &	-0.976 \\
5p $^3$P$^o_1$ -	6d $^3$D$_2$ &	4192.70 &	23850.60$^{vac}$ &	2.50 $\times$	10$^5$ &	-0.499 \\
5p $^3$P$^o_2$ -	6d $^3$D$_1$ &	4190.15 &	23865.51$^{vac}$ &	5.56 $\times$	10$^3$ &	-2.152 \\
5p $^3$P$^o_2$ -	6d $^3$D$_2$ &	4190.13 &	23865.63$^{vac}$ &	8.34 $\times$	10$^4$ &	-0.976 \\
5p $^3$P$^o_2$ -	6d $^3$D$_3$ &	4190.11 &	23865.68$^{vac}$ &	4.67 $\times$	10$^5$ &	-0.228 \\
4d $^1$D$_2$ - 	6p $^1$P$^o_1$ &	4080.37 &	24507.70$^{vac}$ &	2.16 $\times$	10$^5$ &	-1.229 \\
4p $^3$P$^o_0$ -	5s $^3$S$_1$ & 	4031.39 &	24805.24$^{vac}$ &	1.01 $\times$	10$^6$ &	-0.561 \\
4p $^3$P$^o_1$ -	5s $^3$S$_1$ & 	4028.09 &	24825.53$^{vac}$ &	3.03 $\times$    10$^6$ &	-0.084 \\
4p $^3$P$^o_2$ -	5s $^3$S$_1$ & 	4021.34 &	24867.18$^{vac}$ &	5.05 $\times$	10$^6$ &	0.138 \\
4f $^1$F$^o_3$ - 	6g $^1$G$_4$ &	3934.36 &	25417.11$^{vac}$ &	1.33 $\times$	10$^6$ &	0.062 \\
4f $^3$F$^o_3$ -	6g $^3$G$_4$ &	3934.36 &	25417.11$^{vac}$ &	1.24 $\times$	10$^6$ &	-1.376 \\
4f $^1$F$^o_3$ - 	6g $^3$G$_4$ &	3934.36 &	25417.11$^{vac}$ &	1.89 $\times$	10$^5$ &	-0.786 \\
4f $^1$F$^o_3$ - 	6g $^3$G$_3$ &	3934.36 &	25417.11$^{vac}$ &	2.18 $\times$	10$^3$ &	-2.832 \\
4f $^3$F$^o_2$ -	6g $^3$G$_3$ &	3934.15 &	25418.51$^{vac}$ &	1.40 $\times$	10$^6$ &	-0.026 \\
4f $^3$F$^o_3$ -	6g $^3$G$_3$ &	3934.09 &	25418.81$^{vac}$ &	1.20 $\times$	10$^5$ &	-1.092 \\
4f $^3$F$^o_3$ -	6g $^1$G$_4$ &	3934.09 &	25418.81$^{vac}$ &	1.86 $\times$	10$^5$ &	-0.793 \\
4f $^3$F$^o_4$ -	6g $^3$G$_5$ &	3934.05 &	25419.16$^{vac}$ &	1.52 $\times$	10$^6$ &	0.208 \\
4f $^3$F$^o_4$ -	6g $^3$G$_3$ &	3934.04 &	25419.16$^{vac}$ &	1.94 $\times$	10$^3$ &	-2.883 \\
4f $^3$F$^o_4$ -	6g $^3$G$_4$ &	3934.04 &	25419.16$^{vac}$ &	9.01 $\times$	10$^4$ &	-1.107 \\
4f $^3$F$^o_4$ - 	6g $^1$G$_4$ &	3934.04 &	25419.16$^{vac}$ &	4.98 $\times$	10$^3$ &	-2.365 \\
4p $^1$P$^o_1$ -   	4d $^1$D$_2$ & 	3787.88 &	26399.76$^{vac}$ &	5.45 $\times$    10$^6$ &	0.444 \\

4f $^1$F$^o_3$ - 	6d $^3$D$_2$ &	3776.42 &	26480.14$^{vac}$ &	1.04 $\times$	10$^3$ &	-3.262 \\
4f $^1$F$^o_3$ - 	6d $^3$D$_3$ &	3766.40 &	26550.52$^{vac}$ &	9.34 $\times$	10$^1$ &	-4.164 \\
4f $^3$F$^o_2$ -	6d $^3$D$_1$ &	3766.22 &	26551.82$^{vac}$ &	6.61 $\times$	10$^4$ &	-1.682 \\
4f $^3$F$^o_2$ -	6d $^3$D$_2$ &	3766.20 &	26551.97$^{vac}$ &	3.06 $\times$	10$^3$ &	-2.414 \\
4f $^3$F$^o_2$ -	6d $^3$D$_3$ &	3766.19 &	26552.04$^{vac}$ &	1.50 $\times$	10$^2$ &	-3.958 \\
4f $^3$F$^o_3$ -	6d $^3$D$_2$ &	3766.15 &	26552.30$^{vac}$ &	5.78 $\times$	10$^4$ &	-1.519 \\
4f $^3$F$^o_3$ -	6d $^3$D$_3$ &	3766.14 &	26552.37$^{vac}$ &	5.16 $\times$	10$^3$ &	-2.422 \\
4f $^3$F$^o_4$ -	6d $^3$D$_3$ &	3766.09 &	26552.75$^{vac}$ &	6.08 $\times$	10$^4$ &	-1.350 \\
5p $^3$P$^o_0$ -	7s $^3$S$_1$ &  	3606.41 &	27728.44$^{vac}$ &	7.57 $\times$	10$^4$ &	-1.586 \\
5p $^3$P$^o_1$ -	7s $^3$S$_1$ &  	3605.13 &	27738.27$^{vac}$ &	2.27 $\times$	10$^5$ &	-1.109 \\
5p $^3$P$^o_2$ -	7s $^3$S$_1$ &  	3602.49 &	29879.22$^{vac}$ &	3.78 $\times$	10$^5$ &	-0.887 \\
4f $^1$F$^o_3$ - 	6d $^1$D$_2$ &	3346.80 &	29879.22$^{vac}$ &	1.11 $\times$	10$^5$ &	-1.128 \\
4f $^1$F$^o_3$ - 	6d $^1$D$_2$ &	3346.80 &	29879.29$^{vac}$ &	1.11 $\times$	10$^5$ &	-1.128 \\
4f $^3$F$^o_3$ -	6d $^1$D$_2$ &	3346.54 &	29881.57$^{vac}$ &	1.00 $\times$	10$^3$ &	-3.406 \\
5p $^1$P$^o_1$ -  	6d $^1$D$_2$ &	3316.69 &	30150.36$^{vac}$ &	7.69 $\times$	10$^4$ &	-1.293 \\
5p $^1$P$^o_1$ -  	7s $^1$S$_0$ &  	3302.82 &	30277.19$^{vac}$ &	6.21 $\times$	10$^5$ &	-1.065 \\
4p $^1$P$^o_1$ -  	5s $^1$S$_0$ &  	3209.45 &	31157.72$^{vac}$ &	6.11 $\times$	10$^6$ &	-0.052 \\
3d $^1$D$_2$   -	4p $^1$P$^o_1$ &	2943.70 &	33971.27$^{vac}$ &	1.26 $\times$    10$^6$ &	-0.161 \\

4d $^3$D$_3$ -	6p $^3$P$^o_2$ &	2826.79 &	35376.08$^{vac}$ &	7.87 $\times$ 	10$^4$ &	-0.869 \\
4d $^3$D$_2$ -	6p $^3$P$^o_2$ &	2826.73 &	35376.55$^{vac}$ &	1.40 $\times$ 	10$^4$ &	-1.617 \\
4d $^3$D$_1$ -	6p $^3$P$^o_2$ &	2826.69 &	35377.07$^{vac}$ &	9.37 $\times$ 	10$^2$ &	-2.793 \\
4d $^3$D$_2$ -	6p $^3$P$^o_1$ &	2825.47 &	35392.84$^{vac}$ &	4.21 $\times$ 	10$^4$ &	-1.140 \\
4d $^3$D$_1$ -	6p $^3$P$^o_1$ &	2825.39 &	35393.36$^{vac}$ &	1.40 $\times$ 	10$^4$ &	-1.617 \\
4d $^3$D$_1$ -	6p $^3$P$^o_0$ &	2824.74 &	35401.45$^{vac}$ &	1.87 $\times$ 	10$^4$ &	-1.492 \\
5p $^3$P$^o_0$ -	5d $^3$D$_1$ &	2719.43 &	36771.98$^{vac}$ &	2.10 $\times$ 	10$^5$ &	-1.344 \\
5p $^3$P$^o_1$ -	5d $^3$D$_1$ &	2718.19 &	36789.25$^{vac}$ &	1.58 $\times$ 	10$^5$ &	-1.469 \\
5p $^3$P$^o_1$ -	5d $^3$D$_2$ &	2718.12 &	36789.57$^{vac}$ &	4.73 $\times$ 	10$^5$ &	-0.992 \\
5p $^3$P$^o_2$ -	5d $^3$D$_1$ &	2715.55 &	36825.02$^{vac}$ &	1.05 $\times$ 	10$^4$ &	-2.645 \\
5p $^3$P$^o_2$ -	5d $^3$D$_2$ &	2715.52 &	36825.33$^{vac}$ &	1.58 $\times$ 	10$^5$ &	-1.469 \\
5p $^3$P$^o_2$ -	5d $^3$D$_3$ &	2715.45 &	36825.74$^{vac}$ &	8.82 $\times$ 	10$^5$ &	-0.721 \\
4f $^1$F$^o_3$ - 	5g $^1$G$_4$ & 	2586.33 &	38664.95$^{vac}$ &	4.39 $\times$ 	10$^6$ &	0.944 \\
4f $^1$F$^o_3$ - 	5g $^3$G$_4$ &	2586.33 &	38664.95$^{vac}$ &	9.43 $\times$ 	10$^4$ &	-0.724 \\
4f $^1$F$^o_3$ - 	5g $^3$G$_3$ &	2586.32 &	38664.95$^{vac}$ &	6.44 $\times$ 	10$^3$ &	-1.999 \\
4f $^3$F$^o_2$ -	5g $^3$G$_3$ &	2586.11 &	38668.18$^{vac}$ &	4.12 $\times$ 	10$^6$ &	0.807 \\
4f $^3$F$^o_3$ -	5g $^3$G$_3$ &	2586.07 &	38668.88$^{vac}$ &	3.54 $\times$ 	10$^5$ &	-0.259 \\
4f $^3$F$^o_3$ -	5g $^1$G$_4$ &	2586.06 &	38664.95$^{vac}$ &	9.93 $\times$ 	10$^4$ &	-0.702 \\
4f $^3$F$^o_3$ -	5g $^3$G$_4$ &	2586.06 &	38668.88$^{vac}$ &	4.11 $\times$    10$^6$ &	0.916 \\
4f $^3$F$^o_4$ -	5g $^3$G$_3$ &	2586.02 &	38669.69$^{vac}$ &	5.72 $\times$ 	10$^3$ &	-2.050 \\

4f $^3$F$^o_4$ -	5g $^3$G$_4$ &	2586.01 &	38669.69$^{vac}$ &	2.80 $\times$ 	10$^5$ &	-0.251 \\
4f $^3$F$^o_4$ - 	5g $^1$G$_4$ & 	2586.01 &	38669.69$^{vac}$ &	7.20 $\times$ 	10$^1$ &	-3.842 \\
6s $^3$S$_1$ -	7p $^3$P$^o_2$ &	2585.96 &	38670.36$^{vac}$ &	6.30 $\times$ 	10$^4$ &	-0.894 \\
6s $^3$S$_1$ -	7p $^3$P$^o_1$ &	2585.22 &	38681.43$^{vac}$ &	3.78 $\times$ 	10$^4$ &	-1.115 \\
6s $^3$S$_1$ -	7p $^3$P$^o_0$ &	2584.89 &	38686.38$^{vac}$ &	1.26 $\times$ 	10$^4$ &	-1.593 \\
6s $^1$S$_0$ - 	7p $^1$P$^o_1$ & 	2393.36 &	41782.32$^{vac}$ &	8.74 $\times$ 	10$^4$ &	-1.172 \\
5s $^3$S$_1$ - 	5p $^3$P$^o_2$ &	2380.24 &	42013.28$^{vac}$ &	1.21 $\times$ 	10$^6$ &	0.463 \\
5s $^3$S$_1$ - 	5p $^3$P$^o_1$ &	2377.60 &	42059.93$^{vac}$ &	7.24 $\times$ 	10$^5$ &	0.241 \\
5s $^3$S$_1$ - 	5p $^3$P$^o_0$ & 	2376.31 &	42082.53$^{vac}$ &	2.41 $\times$ 	10$^5$ &	-0.236 \\
6p $^3$P$^o_0$ -	7d $^3$D$_1$ &	2301.72 &	43445.87$^{vac}$ &	3.53 $\times$ 	10$^4$ &	-0.828 \\
6p $^3$P$^o_1$ -	7d $^3$D$_1$ &	2301.07 &	43458.06$^{vac}$ &	2.65 $\times$ 	10$^4$ &	-0.953 \\
6p $^3$P$^o_1$ -	7d $^3$D$_2$ &	2301.05 &	43458.40$^{vac}$ &	7.95 $\times$ 	10$^4$ &	-0.476 \\
6p $^3$P$^o_2$ -	7d $^3$D$_1$ &	2299.77 &	43482.65$^{vac}$ &	1.77 $\times$ 	10$^3$ &	-2.129 \\
6p $^3$P$^o_2$ - 	7d $^3$D$_2$ &	2299.75 &	43482.99$^{vac}$ &	2.65 $\times$ 	10$^4$ &	-0.953 \\
6p $^3$P$^o_2$ - 	7d $^3$D$_3$ &	2299.74 &	43483.00$^{vac}$ &	1.48 $\times$ 	10$^5$ &	-0.205 \\
4f $^1$F$^o_3$ - 	5d $^3$D$_2$ &	2291.81 &	43633.65$^{vac}$ &	2.50 $\times$ 	10$^3$ &	-2.453 \\
4f $^1$F$^o_3$ - 	5d $^3$D$_3$ &	2291.78 &	43634.22$^{vac}$ &	2.24 $\times$ 	10$^2$ &	-3.354 \\

4f $^3$F$^o_2$ -	5d $^3$D$_1$ &	2291.62 &	43637.31$^{vac}$ &	1.59 $\times$	10$^5$ &	-0.872 \\
4f $^3$F$^o_2$ -	5d $^3$D$_2$ &	2291.59 &	43637.75$^{vac}$ &	1.76 $\times$	10$^4$ &	-1.604 \\
4f $^3$F$^o_2$ -	5d $^3$D$_3$ &	2291.56 &	43638.00$^{vac}$ &	3.59 $\times$	10$^2$ &	-3.148 \\
4f $^3$F$^o_3$ -	5d $^3$D$_3$ &	2291.52 &	43639.21$^{vac}$ &	1.24 $\times$	10$^4$ &	-1.612 \\
4f $^3$F$^o_4$ -	5d $^3$D$_3$ &	2291.50 &	43630.24$^{vac}$ &	1.46 $\times$	10$^5$ &	-0.540 \\
4f $^3$F$^o_3$ -	5d $^3$D$_2$ &	2291.50 &	43648.64$^{vac}$ &	1.38 $\times$	10$^5$ &	-0.709 \\
5d $^1$D$_2$ - 	7p $^1$P$^o_1$ & 	2271.86 &	44016.80$^{vac}$ &	1.01 $\times$	10$^5$ &	-1.046 \\
5s $^1$S$_0$ -	5p $^1$P$^o_1$ &	2150.35 &	46503.99$^{vac}$ &	1.80 $\times$	10$^6$ &	0.237 \\
6p $^1$P$^o_1$ -  	7d $^1$D$_2$ &	1826.03 &	54763.70$^{vac}$ &	4.18 $\times$	10$^4$ &	-1.050 \\
5p $^3$P$^o_0$ - 	6s $^3$S$_1$ &  	1642.99 &	60864.61$^{vac}$ &	2.42 $\times$	10$^5$ &	-0.401 \\
5p $^3$P$^o_1$ -	6s $^3$S$_1$ &  	1641.71 &	60911.95$^{vac}$ &	7.26 $\times$	10$^5$ &	0.076 \\
5p $^3$P$^o_2$ -	6s $^3$S$_1$ &  	1639.07 &	61010.06$^{vac}$ &	1.21 $\times$	10$^6$ &	0.298 \\
4f $^3$F$^o_3$ -	5d $^1$D$_2$ &	1631.94 &	61276.65$^{vac}$ &	5.81 $\times$	10$^3$ &	-1.762 \\
4f $^1$F$^o_3$ - 	5d $^1$D$_2$ & 	1631.94 &	61276.65$^{vac}$ &	3.22 $\times$	10$^5$ &	-0.020 \\
5p $^1$P$^o_1$ -  	5d $^1$D$_2$ & 	1601.85 &	62428.01$^{vac}$ &	1.38 $\times$	10$^6$ &	0.587 \\
4d $^1$D$_2$ -	5p $^1$P$^o_1$ &	1571.89 &	63617.52$^{vac}$ &	7.46 $\times$	10$^5$ &	0.147 \\
4d $^1$D$_2$ -	4f $^3$F$^o_3$ &	1542.06 &	64848.36$^{vac}$ &	1.64 $\times$	10$^4$ &	-1.206 \\
4d $^1$D$_2$ - 	4f $^1$F$^o_3$ &	1541.80 &	64859.42$^{vac}$ &	8.30 $\times$	10$^5$ &	0.537 \\
5d $^3$D$_3$ -	7p $^3$P$^o_2$ &	1509.54 &	66245.26$^{vac}$ &	3.69 $\times$	10$^4$ &	-0.651 \\
5d $^3$D$_2$ -	7p $^3$P$^o_2$ &	1509.51 &	66246.58$^{vac}$ &	6.60 $\times$	10$^3$ &	-1.400 \\
5d $^3$D$_1$ - 	7p $^3$P$^o_2$ &	1509.49 &	66247.58$^{vac}$ &	4.40 $\times$	10$^2$ &	-2.576 \\

5d $^3$D$_2$ - 	7p $^3$P$^o_1$ &	1508.77 &	66279.07$^{vac}$ &	1.98 $\times$	10$^4$ &	-0.922 \\
5d $^3$D$_1$ - 	7p $^3$P$^o_1$ &	1508.75 &	66280.08$^{vac}$ &	6.60 $\times$	10$^3$ &	-1.400 \\
5d $^3$D$_1$ - 	7p $^3$P$^o_0$ &	1508.42 &	66294.62$^{vac}$ &	8.80 $\times$	10$^3$ &	-1.275 \\
5p $^1$P$^o_1$ -  	6s $^1$S$_0$ &  	1480.34 &	67552.19$^{vac}$ &	1.84 $\times$	10$^6$ &	0.108 \\
6p $^3$P$^o_0$ -	6d $^3$D$_1$ &	1425.80 &	70136.26$^{vac}$ &	6.13 $\times$	10$^4$ &	-0.176 \\
6p $^3$P$^o_1$ -	6d $^3$D$_1$ &	1425.15 &	70168.05$^{vac}$ &	4.60 $\times$	10$^4$ &	-0.301 \\
6p $^3$P$^o_1$ -	6d $^3$D$_2$ &	1425.13 &	70169.09$^{vac}$ &	1.38 $\times$	10$^5$ &	0.176 \\
6p $^3$P$^o_2$ -	6d $^3$D$_1$ &	1423.85 &	70232.17$^{vac}$ &	3.07 $\times$	10$^3$ &	-1.477 \\
6p $^3$P$^o_2$ -	6d $^3$D$_2$ &	1423.83 &	70233.20$^{vac}$&	4.60 $\times$	10$^4$ &	-0.301 \\
6p $^3$P$^o_2$ -	6d $^3$D$_3$ &	1423.82 &	70233.70$^{vac}$ &	2.57 $\times$	10$^5$ &	0.447 \\
6s $^3$S$_1$ -	6p $^3$P$^o_2$ &	1127.25 &	88713.40$^{vac}$&	3.28 $\times$	10$^5$ &	0.548 \\
6s $^3$S$_1$ -	6p $^3$P$^o_1$ &	1125.93 &	88815.90$^{vac}$&	1.97 $\times$	10$^5$ &	0.326 \\
6s $^3$S$_1$ -	6p $^3$P$^o_0$ &	1125.29 &	88866.90$^{vac}$ &	6.56 $\times$	10$^4$ &	-0.151 \\
6s $^1$S$_0$ - 	6p $^1$P$^o_1$ &	1028.12 &	97265.01$^{vac}$ &	5.20 $\times$	10$^5$ &	0.328 \\

\hline

\end{longtable} 
\end{longtab}

\begin{acknowledgements}
We acknowledge the grant no $621-2011-4206$ from the Swedish Research Council (VR) and Crafoord foundation grant 2015-0947 . The infrared FTS at the Edl{\'e}n laboratory is made available through a grant from the Knut and Alice Wallenberg Foundation. We are grateful to Hans Lundberg for revisiting the laser measurements. A.P.R. acknowledges the travel grant for young researches from the Royal Physiographic Society of Lund. We are grateful for discussions with Paul Barklem, Nils Ryde, and Henrik J{\"o}nsson. This project was supported by `The New Milky Way' project from the Knut and Alice Wallenberg foundation.
     \end{acknowledgements}


\bibliography{bibli} 
\bibliographystyle{aa}

\end{document}